\RequirePackage{lineno}
\documentclass[superscriptaddress,preprint,amsmath,amssymb,prb]{revtex4}

\usepackage{graphicx}
\usepackage{dcolumn}
\usepackage{bm}
\usepackage{color,soul}

\usepackage{hyperref}

\usepackage{tabularx}
\usepackage{float}
\usepackage{booktabs} 
\usepackage{subfigure} 
\usepackage{longtable}%
\usepackage[resetlabels]{multibib}

\newcommand{\SBP}[1]{\color{black}{#1} \color{black}}

\begin{document}

\title{Kondo-like phonon scattering in thermoelectric clathrates}
\author{M.~Ikeda}
\affiliation{Institute of Solid State Physics, Vienna University of Technology, Wiedner Hauptstr.\ 8-10, 1040 Vienna, Austria}
\author{H.~Euchner}
\affiliation{Institute of Materials Science and Technology, Vienna University of Technology, Getreidemarkt 9, 1060 Vienna, Austria}
\affiliation{Institute of Theoretical Chemistry, Ulm University,
Albert-Einstein-Allee 11, 89069 Ulm, Germany}
\author{X.~Yan}
\author{P.~Tome\v{s}}
\author{A.~Prokofiev}
\author{L.~Prochaska}
\affiliation{Institute of Solid State Physics, Vienna University of Technology, Wiedner Hauptstr.\ 8-10, 1040 Vienna, Austria}
\author{G.~Lientschnig}
\affiliation{Institute of Solid State Physics, Vienna University of Technology, Wiedner Hauptstr.\ 8-10, 1040 Vienna, Austria}
\affiliation{Center for Micro- and Nanostructures, Vienna University of Technology, Floragasse 7, 1040 Vienna, Austria}
\author{R.~Svagera}
\affiliation{Institute of Solid State Physics, Vienna University of Technology, Wiedner Hauptstr.\ 8-10, 1040 Vienna, Austria}
\author{S.~Hartmann}
\author{E.~Gati}
\author{M.~Lang}
\affiliation{Physikalisches Institut, Goethe-Universit\"at Frankfurt, Max-von-Laue-Stra{\ss}e 1, 60438 Frankfurt am Main, Germany}
\author{S.~Paschen$^{\ast}$}
\affiliation{Institute of Solid State Physics, Vienna University of Technology, Wiedner Hauptstr.\ 8-10, 1040 Vienna, Austria}
\date{September 29, 2017}
\maketitle

\noindent{\bf Crystalline solids are generally known as excellent heat
conductors, amorphous materials or glasses as thermal insulators. It has thus
come as a surprise that certain crystal structures defy this paradigm. A
prominent example are type-I clathrates and \SBP{other} materials with
guest-host structures. They sustain low-energy Einstein-like modes in their
phonon spectra, but are also prone to various types of disorder and
phonon-electron scattering \SBP{and thus the mechanism} responsible for their
ultralow thermal conductivities has remained elusive. \SBP{While recent {\em ab
initio} lattice dynamics simulations show that the Einstein-like modes enhance
phonon-phonon Umklapp scattering, they reproduce experimental data only at low
temperatures. Here we show that a new effect, an ``all phononic Kondo effect'',
can resolve this discrepancy. This is evidenced by our thermodynamic and
transport measurements on various clathrate single crystal series and their
comparison with {\em ab initio} simulations. Our new understanding devises
design strategies to further suppress the thermal conductivity of clathrates and
other related materials classes, with relevance for the field of thermoelectric
waste heat recovery but also more generally for phononic applications. More
fundamentally, it may trigger theoretical work on strong correlation effects in
phonon systems.}} \newpage





\noindent{\bf INTRODUCTION}

The increasing world wide energy consumption and the associated climate change
call for enhancing the overall energy efficiency of technological energy
conversion processes.
\SBP{Thermoelectrics\cite{Sla95.1,Row06.2,Nie16.1,Shi16.1,Gor17.1,Hee17.1}} can convert
waste heat into electricity and could thus contribute to such an efficiency
increase. The perfect thermoelectric material combines low phonon thermal
conductivity with high electrical conductivity and high thermopower. Materials
classes with cage-like structures such as the
\SBP{clathrates\cite{Tak14.1,Nol14.1}} or
\SBP{skutterudites\cite{Lee06.1,Shi16.1}} show surprisingly low phonon thermal
conductivities $\kappa_{\rm ph}$, even in perfect single crystals. Type-I
clathrates \SBP{G$_8$H$_{46}$,} that show promising thermoelectric figures of
\SBP{merit\cite{Sar06.1,Tak13.1,Gor17.1},} consist of host atoms (H) that
encapsulate guest atoms (G) in a framework of oversized cages
(Fig.\,\ref{fig:kappa2}a).
Recent inelastic neutron and X-ray scattering studies uncovered energetically
low-lying optical phonon modes which were attributed to a ``rattling'' motion of
the guest atoms in the cages and are thus referred to as Einstein
modes\cite{Chr08.1,Euc12.1,Pai14.1}. \SBP{Similar observations were also made
for skutterudites\cite{Lee06.1} and clathrate hydrates\cite{Tse01.1}. A severe
flattening of the acoustic phonon branches at energies near the optical modes
was observed\cite{Chr08.1,Euc12.1,Pai14.1}, and attributed to a finite coupling
between the guest atoms and the host cages. Early thermal conductivity studies
on type-I clathrates\cite{Coh99.1,Nol00.1}, inspired by investigations on
glasses\cite{Gra86.1}, tried to capture this guest-host coupling by including
terms for resonance scattering, originally proposed to describe the resonance
interaction between phonons and non-paramagnetic defects\cite{Poh62.1}, and for
scattering from tunneling states\cite{Gra86.1} into phenomenological
treatments\cite{Cal59.1}. Whereas empirical multi-parameter fits including these
terms can indeed model thermal conductivity data of several type-I clathrates
below about 50\,K (Refs.\citenum{Coh99.1,Nol00.1}), inconsistencies with other
results have been pointed out more recently. Notably, such modelling can neither
explain the exceptionally long phonon lifetimes\cite{Chr08.1,Lor17.1} nor the
large thermal conductivity differences between structurally very similar n- and
p-type versions of these materials\cite{Chr06.1}. The recent achievement of {\em
ab initio} calculations of the thermal conductivity of a type-I clathrate, based
on intrinsic phonon-phonon Umklapp scattering processes\cite{Tad15.1},
represents a major step forward. It showed that, unlike in the resonance
scattering picture, the phonon lifetimes are reduced in a wide frequency range.
Nevertheless, also these calculations can explain experimental thermal
conductivity data only below about 50\,K. Here, we propose a new mechanism, the
``all phononic Kondo effect'', that extends the agreement all the way up to room
temperaturere. As we will show, the pronounced difference in temperature
dependence is due to the disentanglement of Kondo-coupled acoustic and rattling
phonon modes above the phonon Kondo temperature.}

\SBP{In general terms, the} Kondo effect describes non-commutative scattering of an extended wave from a
localized entity with internal degree of freedom. It has been studied in many
different settings, including magnetic, orbital, charge, and local vibrational
degrees of freedom (see Supplementary Material S8 for more details). However, to
the best of our knowledge, it has never been studied in an all phononic context.

\begin{figure*}[ht!]
{\large\bf \hspace{-7.3cm} a \hspace{7cm} b}
\vspace{-0.5cm}

\centering
\subfigure{\includegraphics[width=0.45\textwidth]{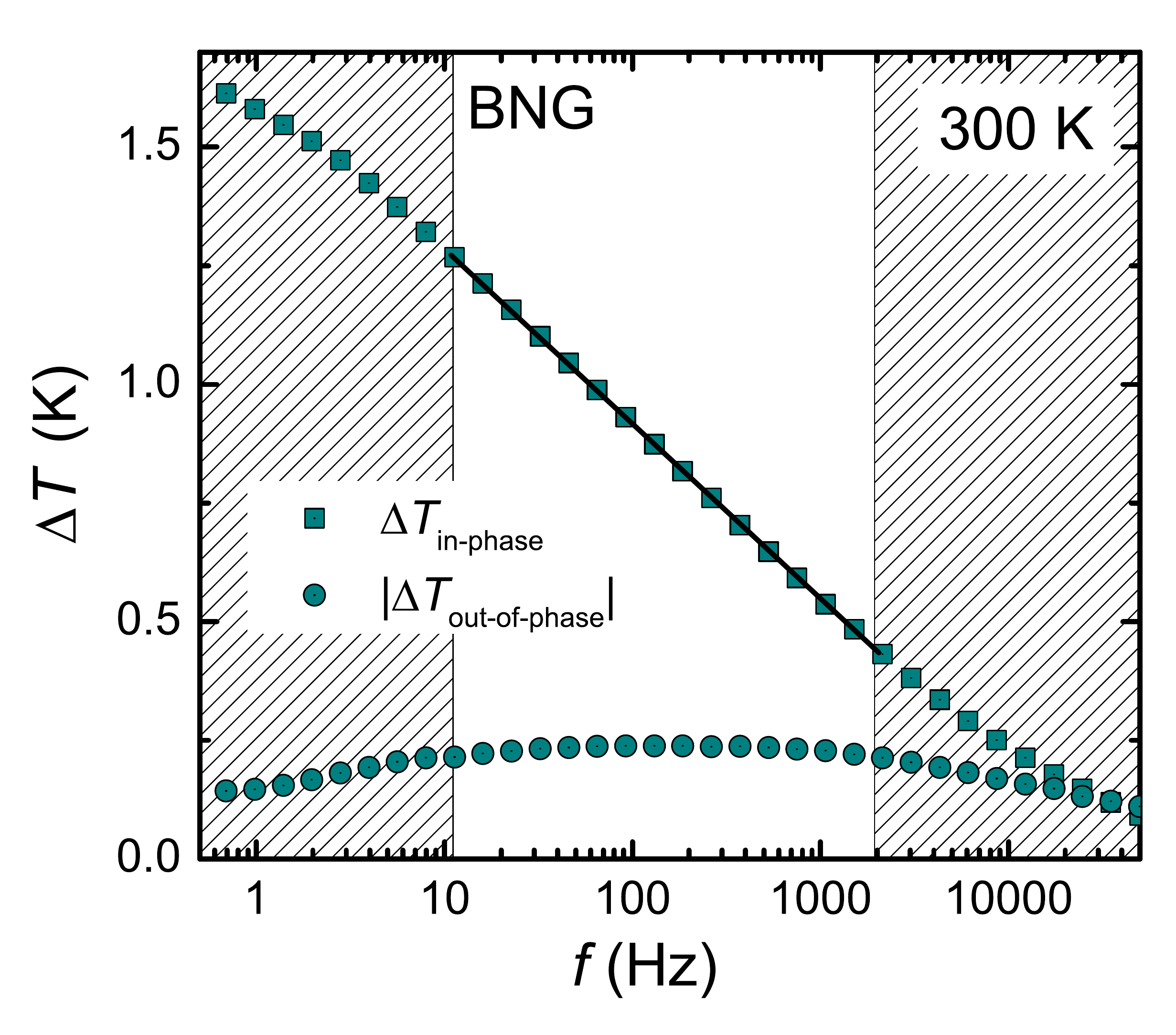}}\hfill
\subfigure{\includegraphics[width=0.51\textwidth]{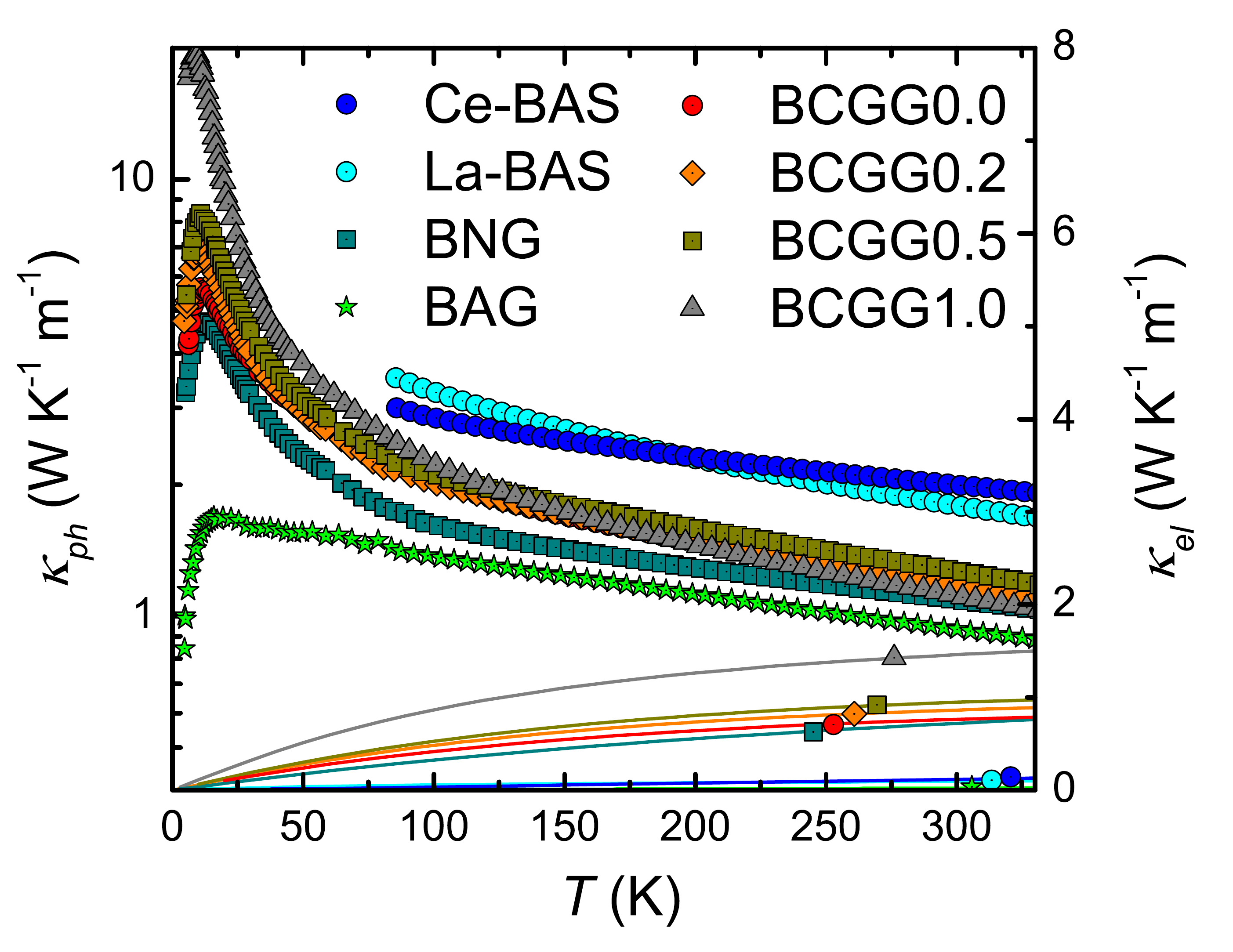}}\\
\caption{{\bf Thermal conductivity data of type-I clathrate single crystals.}
({\bf a}) Exemplary frequency scan in the $3\omega$ setup. Rectangles show the
in-phase, circles the out-of-phase contribution of the temperature oscillation
$\Delta T$. The expected linear-in-logarithmic-frequency dependence of the
in-phase temperature oscillation of the heater/thermometer line, that is
observed at intermediate frequencies (between the hatched areas), is measured
with an extremely low noise level. The slope of this dependence (full line) is
inversely proportional to the thermal conductivity $\kappa$ of the studied
material. ({\bf b}) Temperature dependent thermal conductivity of all studied
materials. The data points represent the phonon thermal conductivities
$\kappa_{\rm ph}$, calculated by subtracting an electronic contribution (lines
and identifier symbols) from the total measured $\kappa$. Each data point above
80\,K is determined from the slope of an isothermal $\Delta T$ vs $\log f$ curve
(straight line in a). All sample compositions are given in the Supplementary
Table~1.}
\label{fig:kappa}
\end{figure*}

To demonstrate this new effect, we had to go beyond the
state-of-the-art in thermoelectric clathrates' research \SBP{in several respects.} Firstly, to overcome
the bottleneck of large uncertainties and systematic errors of standard thermal
conductivity measurements at elevated temperatures, we developed a
high-precision implementation (Fig.\,\ref{fig:kappa}a and Materials and Methods)
of the 3$\omega$ technique\cite{Cah87.1}. A second key ingredient for our work
are sample series of single crystals, which allow us to pin down the dominating
{\em instrinsic} phonon scattering mechanisms \SBP{by employing an empirical
model\cite{Cal59.1}, modified to be in line with the most recent
findings\cite{Tad15.1,Lor17.1} on the phonon band structure and phonon dynamics
in type-I clathrates.} Thirdly, we compare both the thermal conductivity data
and thermodynamic data to {\em ab initio} lattice dynamics calculations, thereby
providing compelling evidence for the phonon Kondo effect.\\[-0.4cm]

\noindent{\bf RESULTS}

We start by giving an overview of the phonon thermal conductivities
$\kappa_{\rm ph}$ of all type-I clathrate single crystals studied here
(Fig.\,\ref{fig:kappa}b). Although type-I clathrates are often referred to as
model systems for the enigmatic phonon-glass electron-crystal
concept\cite{Sla95.1}, the most prominent feature of $\kappa_{\rm ph}(T)$ is a
large crystal-like maximum. It is narrow and occurs at a surprisingly low
temperature of only about 10\,K. As will be further detailed below, this narrow
low-temperature maximum is a strong indication for the phonon transport not
being controlled by the natural energy scale $k_{\rm B}\Theta_{\rm D}$ of the
Debye temperature $\Theta_{\rm D}$ but by \SBP{a much smaller} energy scale
$k_{\rm B}\Theta_{\rm E}$.

To visualize the effect of a reduced Debye temperature we use a simple
phenomenological lattice thermal conductivity model: A modified version of the
standard Callaway model\cite{Cal59.1} where we replaced the Debye temperature
$\Theta_{\rm D}$ by a variable temperature $\Theta_{\rm E}$
(Supplementary Material S1). The left panel of Fig.\,\ref{fig:kappa2}d shows
temperature dependencies of normalized phonon thermal conductivities calculated
within this model for a series of different characteristic energies $k_{\rm
B}\Theta_{\rm E}$, keeping fixed scattering rates $\tau^{-1}_{\rm D}$ for defect
scattering, $\tau^{-1}_{\rm B}$ for boundary scattering, and $\tau^{-1}_{\rm
ph-el}$ for phonon-electron scattering. Resonance scattering is of minor
importance here (Supplementary Fig.\,4). With decreasing $\Theta_{\rm E}$ the
broad maximum sharpens and is shifted to lower temperatures. This is attributed
to a strongly enhanced Umklapp scattering rate $\tau^{-1}_{\rm U}$
(Supplementary Material S1) that results from successively limiting the
energy range of the contributing acoustic phonons as $\Theta_{\rm E}$ decreases.

This effect can be directly observed by experiments. The right panel of
Fig.\,\ref{fig:kappa2}d compares the phonon thermal conductivities of a Ge-based
type-I clathrate single crystal (BCGG1.0, see Supplementary Table~1) and single
crystalline Ge, irradiated\cite{Bry61.1} to a similar defect concentration. Both
materials have similar Debye temperatures (Supplementary Material S2), but in
the clathrate the low-lying Einstein temperature dominates the behaviour.

Next we prove, with two clathrate single crystal series, that the strong
Umklapp scattering due to the new low-energy scale indeed dominates the lattice
thermal conductivity in a wide temperature range. In the first such series,
Ba$_8$Cu$_{4.8}$Ge$_{\rm 41.2-x-y}\square_{\rm y}$Ga$_{\rm x}$ ($x=0, 0.2, 0.5,
1.0$, see BCGG$x$ in Supplementary Table~1), an increasing Ga content $x$ is
accompanied by a decreasing content $y$ of host atom vacancies $\square$. This
is supported by the increase of the lattice parameter $a$
(Fig.\,\ref{fig:scattering}a) 
and the Hall mobility $\mu_{\rm H}=R_{\rm
H}/\rho=1/(ne\rho)$ (Fig.\,\ref{fig:scattering}c) with $x$, as well as by the
charge neutrality of this substitution (both the electronic $\gamma$ term of the
specific heat and the charge carrier concentration $n$ are essentially
independent of $x$, see Fig.\,\ref{fig:scattering}b,\,c bottom), as discussed in
the Supplementary Material S3. Because the mass difference between Ga and Ge
is very small, Ga represents a much weaker scattering potential\cite{Kei15.1} in
Ba$_8$Cu$_{4.8}$Ge$_{\rm 41.2-x-y}\square_{\rm y}$Ga$_{\rm x}$ than a vacancy.
Thus, we expect a decrease of defect scattering with $x$. The low-temperature
phonon thermal conductivity $\kappa_{\rm ph}(T)$ of all samples in this series
shows a maximum at low temperatures that is systematically enhanced at nearly
constant temperature with increasing $x$ (Fig.\,\ref{fig:scattering}d left).
This trend is indeed well reproduced by a decrease of the point-defect
scattering rate $\tau_{\rm D}^{-1}$ with $x$, as seen by the good agreement of
simulation and data for this case (full and dashed lines in
Fig.\,\ref{fig:scattering}d left). A much less satisfying agreement is observed
if the boundary scattering rate $\tau_{\rm B}^{-1}$ is allowed to change with
$x$ instead (Fig.\,\ref{fig:scattering}d right). The phonon-electron scattering
rate $\tau^{-1}_{\rm ph-el}$ can be assumed to be $x$ independent because of the
above discussed charge neutrality. Above about 50\,K, however, the $\kappa_{\rm
ph}(T)$ data for different $x$ merge (Fig.\,\ref{fig:scattering}d left),
indicating that scattering from point defects has become negligible. This can be
rationalized by comparing the (Debye) phonon wavelength $\lambda=2\pi v_{\rm
s}/\omega$, where $v_{\rm s}$ is the sound velocity (Fig.\,\ref{fig:kappa2}c
right), with the size of the scattering center. A vacancy with the distortion
surrounding it was estimated to have a diameter $d \approx 5$\,\AA\
(Ref.\,\onlinecite{Ngu10.1}). Strong point-defect scattering of Rayleigh type,
with an $\omega^4$ dependence, occurs only if $\lambda$ is at least an order of
magnitude larger than $d$, corresponding to phonon energies $\hbar\omega$ of
2.5\,meV (30\,K) and below. At much larger energies (and temperatures), defect
scattering should be weak and frequency independent\cite{Kle51.1}.

The second sample series is the prototypal clathrate Ba$_8$Ga$_{\rm
16-x}$Ge$_{\rm 30+x}$ that has been much investigated in the past. It is an
ideal system to study the importance of phonon-electron scattering because it
has a low and essentially constant amount of point defects (as seen above and shown in Ref.\,\onlinecite{Kei15.1},
Ga does not act as strong point defect in Ge clathrates) but a charge
carrier concentration that varies strongly with $x$
(Ref.\,\onlinecite{May09.1}). Indeed, different Ba$_8$Ga$_{\rm 16-x}$Ge$_{\rm
30+x}$ single crystals reported in the literature\cite{Ben04.1,Avi06.1} show
severely different $\kappa_{\rm ph}(T)$ at low temperatures.
Figure\,\ref{fig:scattering}e replots two extreme cases. The red line is our
low-temperature fit ($T<\Theta_{\rm E}/2$) with the modified Callaway model
(Supplementary Eqn.\,1) to the data of Ref.\,\onlinecite{Ben04.1}. It takes
Umklapp scattering ($\tau^{-1}_{\rm U}$), defect scattering ($\tau^{-1}_{\rm
D}$), boundary scattering ($\tau^{-1}_{\rm B}$), and phonon-electron scattering
($\tau^{-1}_{\rm ph-el}$) into account. The latter is calculated from the
reported materials properties\cite{Ben04.1} using Supplementary Eqn.\,8.
Interestingly, the $\kappa_{\rm ph}(T)$ data of a different Ba$_8$Ga$_{\rm
16-x}$Ge$_{\rm 30+x}$ single crystal\cite{Avi06.1} can be very well reproduced
by strongly increasing $\tau^{-1}_{\rm ph-el}$ and by only slightly adjusting
$\tau^{-1}_{\rm U}$ (dashed line in Fig.\,\ref{fig:scattering}e). Above about
50\,K, the differences in $\kappa_{\rm ph}$ caused by a different
$\tau^{-1}_{\rm ph-el}$ vanish. This observation is in line with the fact that
phonon-electron scattering can only occur for phonons with a wave vector $q$
smaller than twice the Fermi wave vector $k_{\rm F}$. For the crystal of
Ref.\,\onlinecite{Ben04.1} we estimate this to hold below about 140\,K
(Supplementary Material S2). At higher temperatures phonon-electron
scattering is unlikely to be relevant.

A suppression of large low-temperature differences in $\kappa_{\rm ph}(T)$
at higher temperatures has also been seen in other type-I clathrate single
crystal series\cite{Sue07.1,Chr16.1}, but precise high-temperature data on these
are not available to date.

Taking both our clathrate series together we have managed to rule out the
influence of defect and phonon-electron scattering on $\kappa_{\rm ph}(T)$ of
various type-I clathrates above 50\,K. In addition, boundary scattering cannot
contribute significantly in single crystals at these temperatures. Thus, at
elevated temperatures, $\kappa_{\rm ph}$ is dominated by intrinsic phonon-phonon
(Umklapp) scattering. This allows us to pin down its microscopic origin, as
shown in what follows.

Remarkably, a broad range of clathrates, including even a gas
hydrate\cite{Tse97.1,Han87.1,Tse01.1,Gab09.1}, shows a universal scaling of the
room-temperature phonon thermal conductivity with the product of sound velocity
and Einstein temperature of the lowest-lying rattling mode(s), $\kappa_{\rm ph}
\propto v_{\rm s} \Theta_{\rm E}$ (Fig.\,\ref{fig:universal}). The simple
kinetic gas relation $\kappa_{\rm ph} = c_{\rm v} v_{\rm s}^2 \tau/3$ predicts
$\kappa_{\rm ph}$ to depend on the square of the sound velocity. In the Debye
model the sound velocity is proportional to the Debye temperature and thus
$\kappa_{\rm ph} \propto v_{\rm s}^2 \propto \Theta_{\rm D}^2$ is expected for
simple Debye solids. The modified scaling $\kappa_{\rm ph} \propto v_{\rm s}
\Theta_{\rm E} \propto v_{\rm s}^2 \cdot (\Theta_{\rm E}/\Theta_{\rm D})$
shows that the above discussed energy renormalization is universal in
clathrates.

A similar energy renormalization is seen in heavy fermion metals, where the
(spin) Kondo effect rescales the Fermi temperature $T_{\rm{F}}$ to the (spin)
Kondo temperature $T_{\rm{K}}$ (Refs.\,\onlinecite{Ste84.1,Col15.1}).
Figure\,\ref{fig:kappa2}c illustrates this  analogy with schematic dispersion
relations for electrons and phonons. In a band
picture for heavy electron systems (left) a broad conduction band (blue)
hybridizes with a flat, essentially non-dispersing $4f$ band of (renormalized)
energy $\epsilon_{\rm 4f}$ (red). The hybridized bands (violet) are extremely
flat near $\epsilon_{\rm 4f}$, corresponding to quasiparticles with strongly
renormalized effective masses. In the phonon case the strongly dispersing
acoustic phonon mode, approximated here by the linear dispersion of the Debye
model, takes the role of the broad conduction band, and the flat Einstein-like
rattling mode at $\hbar\omega_{\rm E}=k_{\rm B}\Theta_{\rm E}$ corresponds to
the narrow $4f$ band. The resulting hybridized band is again extremely flat in
large portions of the Brillouin zone, giving rise to ``heavy'' phonons with
extremely low group velocity $v_{\rm g}$ at finite wave vectors. \SBP{Even though the resulting $T=0$ dispersion relation may look similar to the one obtained from {\em ab initio} lattice dynamics simulations\cite{Tad15.1}, there is an important difference: Being a strong correlation phenomenon, the phonon Kondo effect has a characteristic temperature dependence. Well above the Kondo temperature, the interacting states (violet) go over to the non-interacting ones (red and blue), an effect referred to as crossover from infrared slavery to asymptotic freedom\cite{Col15.1}, which is absent in the lattice dynamics simulations (see also Discussion).}

\SBP{The fact that the observed universal scaling of $\kappa_{\rm ph}$ still
contains $v_{\rm s}$ (to linear power) is attributed} to the absence of a Fermi
level in phonon systems. Whereas the electrical resistivity in heavy fermion
metals is dominated by the heavy electrons at the Fermi wavevector $k_{\rm F}$,
phonons of all wave vectors, including long-wavelength ones propagating with
$v_{\rm s}$, contribute to $\kappa_{\rm ph}$.

Finally, we compare specific heat, thermal expansion, and thermal conductivity
data for the prototypal clathrate Ba$_8$Ga$_{16}$Ge$_{30}$ with {\em ab initio}
lattice dynamics calculations (see Materials and Methods) and reveal key
characteristics of the Kondo effect. For the specific heat, the agreement
between experiment and theory is excellent at low and high temperatures, but a
distinct deviation is seen in between (Supplementary Fig.\,1b). This difference
gives rise to an ``anomaly'' near $\Theta_E$ (Fig.\,\ref{fig:kappa2}e). It
releases an entropy of order $R\ln 2$ per rattling atom at the $6d$ site. This
is the behaviour expected for any Kondo effect involving a 2-fold degenerate
localized entity: Its degeneracy is lifted by the Kondo interaction, giving rise
to the above entropy release. The entropy reaches $0.4 \ln{2}$ at 28\,K. Twice
this temperature is generally considered as good estimate of the Kondo
temperature\cite{Mel82.1,Geg08.1}, which is thus very close to $\Theta_{\rm E}$.
The experimental thermal expansion is sizeably smaller than the theoretical
prediction below 150\,K, but agrees well with it above this temperature
(Supplementary Fig.\,2b). The difference between the theoretical and
experimental curve closely resembles the specific heat anomaly
(Fig.\,\ref{fig:kappa2}e). \SBP{These discrepancies in specific heat and thermal
expansion translate into a corresponding discrepancy in the Gr\"{u}neisen
parameter: {\em Ab initio} calculations predict an upturn at low temperatures,
an effect usually associated with enhanced phonon anharmonicity at low
frequencies, that is however absent in the experimental curve (Supplementary
Fig.\,S2c). Finally, the} experimental phonon thermal conductivity is well
described by the \SBP{{\em ab initio}} calculations below 50\,K, but severely
overshoots them at higher temperatures (Supplementary Fig.\,3b). The difference,
plotted as inverse to represent a thermal resistivity (Fig.\,\ref{fig:kappa2}f),
{\em in}creases with {\em de}creasing temperature, with a slow (about $-\ln{T}$)
dependence, the hallmark of incoherent Kondo scattering above $T_{\rm K}$ in the
original spin Kondo effect\cite{Ste84.1,Col15.1}.\\[-0.4cm]

\noindent{\bf DISCUSSION}

To understand these results it is important to \SBP{assess the limitations of}
the used {\em ab initio} calculations (see Materials and Methods). The phonon
density of states (DOS) and the specific heat are calculated in the harmonic
approximation, the thermal expansion in the quasi-harmonic approximation, and
the thermal conductivity from anharmonic interatomic force
constants\cite{Tad15.1}. \SBP{The latter thus takes phonon-phonon interactions
to lowest order into account. However, temperature dependencies resulting from
strong phonon correlation effects, most notably the Kondo disentanglement of
acoustic and rattling modes above the phonon Kondo temperature, cannot be
captured by these simulations.} This is why the comparison of experimental data
and such calculations can be used to extract the characteristic temperature
dependencies due to the correlations (somewhat like non-$f$ reference materials
are used as background to reveal Kondo physics in heavy electron compounds --
unfortunately, ``empty'' type-I clathrates without the rattling atoms, that
would represent such background here, do not exist). Specifically, the
temperature-dependent specific heat is overestimated by the calculations as,
unlike in the Kondo effect, no entropy is released by Bose populating
temperature-independent anticrossings (circles in Fig.\,\ref{fig:kappa2}e). The
thermal expansion calculations do not contain the non-trivial temperature
dependence of the anharmonicities due to the phonon Kondo interaction, which
leaves a corresponding imprint in the thermal expansion difference (stars in
Fig.\,\ref{fig:kappa2}e). Finally, the strong Umklapp scattering that is
captured by the thermal conductivity calculations does not contain the weakening
above $T_{\rm K}$ characteristic of an asymptotically free theory
(Fig.\,\ref{fig:kappa2}f).

Further support, independent of any theoretical modeling, comes from recent
inelastic neutron scattering experiments. They reveal unexpected changes of the
optical dynamical structure factors as function of temperature\cite{Lor17.1}
that could be an indication of such a disentanglement of interacting acoustic and rattling modes at high temperatures. All these observations provide strong
evidence for a new correlation effect -- an all phononic Kondo effect -- as
microscopic origin of the peculiar thermodynamic and thermal transport behaviour
of clathrates. 

We do not want to conclude without proposing a possible microscopic
realization of the phonon Kondo effect in clathrates. The rattling motion of the
guest atoms at the crystallographic $6d$ site within a soft
plane\cite{Avi06.1,Lee07.1,Chr08.1} in the tetracaidecahedra 
(Fig.\,\ref{fig:kappa2}b) is known to have the lowest
frequency\cite{Tad15.1}. Within this plane, due to the four-fold symmetry of the
potential well, rattling occurs preferentially along two perpendicular soft
directions\cite{Sal01.1,Chr06.1,Chr07.1}, that we refer to as ``1'' and ``2''.
The two corresponding rattling modes ($e_1$ and $e_2$ in
Fig.\,\ref{fig:kappa2}b), which are degenerate in energy for symmetry reasons,
are thus identified as the pseudospins of the Kondo model (they represent ``spin
up'' and ``spin down'' in the spin Kondo effect). They can be approximated as
simple Einstein oscillators, with the first excited state corresponding to the
``Einstein temperature'' $\Theta_{\rm{E}}$ observed in inelastic scattering
experiments\cite{Chr08.1,Euc12.1,Pai14.1}. The two polarizations of the
transverse acoustic phonons represent the corresponding degree of freedom of the
itinerant species ($e_{\rm{T1}}$ and $e_{\rm{T2}}$ in Fig.\,\ref{fig:kappa2}b,
thus representing the conduction electrons in the spin Kondo effect). The
hybridization of the local (rattling) and extended (acoustic) phonon modes has
been observed in inelastic neutron and X-ray scattering
experiments\cite{Tse97.1,Tse01.1, Chr08.1,Euc12.1,Pai14.1}. A spin flip process
in the spin Kondo effect corresponds to a scattering process with polarization
change in the phonon Kondo effect. It can be visualized as follows: Assume the
guest atom rattles in mode $e_1$ and a propagating phonon of polarization
$e_{\rm{T2}}$ ``hops'' onto the cage, creating a distortion of the cage that
resembles the effect of mode $e_2$. This additional cage distortion will
facilitate a change of rattling direction from $e_1$ to $e_2$, and an
accompanying change in polarization of the outgoing acoustic wave from
$e_{\rm{T2}}$ to $e_{\rm{T1}}$. Such an interaction of two modes with
``opposite'' polarization is analogous to an antiferromagnetic exchange
interaction in the spin Kondo effect. Further details as well as a suggestion
for a mapping of these ingredients onto a Kondo-type Hamiltonian are given in
the Supplementary Material S8.

Our discovery of Kondo physics in an all phononic system is not only of
fundamental interest (see discussion of non-canonical Kondo physics in other
settings in Supplementary Material S8), but also has practical implications.
First, it gives a concise description of how to tailor ``rattling
materials'' for thermoelectric applications at elevated temperatures, which are
most relevant for waste heat-recovery applications: Materials with the lowest
possible Einstein and thus phonon Kondo temperatures should be found.
Microscopically, this might be achieved by changing the mass and size of the
guest atoms, but also by structural disorder on the guest
site\cite{He14.1,Sui15.1} and/or tailored guest-host charge
transfer\cite{He14.1}.

An equally striking consequence is that Cahill's definition of the minimum
thermal conductivity $\kappa_{\rm ph}^{\rm min}$ (Ref.\,\onlinecite{Cah92.1}),
that is being referred to so extensively, needs to be reconsidered. With the
energy renormalization also $\kappa_{\rm ph}^{\rm min}$ is drastically reduced
(Supplementary Material S1). This implies that for the materials in
Fig.\,\ref{fig:universal} there is still room for further improvement as their
phonon thermal conductivities all lie well above their respective new
$\kappa_{\rm ph}^{\rm min}$ value (dashed red line in
Fig.\,\ref{fig:universal}). Further lowering the $\kappa_{\rm ph}$ of a given
material could, for instance, be realized by nanostructuring or the introduction
of dense dislocation arrays\cite{Il15.1}. Long-wavelength phonons that are only
weakly affected by the phonon Kondo effect could, at high temperatures, be
effectively scattered by nanostructures that are small compared to their mean
free path. Indeed, a few clathrates and skutterudites appear to violate the
original Cahill $\kappa_{\rm ph}^{\rm min}$ limit, though this has not been
explicitly recognized in these works (Supplementary Material S2). We expect
our results to trigger more systematic efforts along these lines.

The occurrence of low-lying Einstein-like phonon modes that interact with
acoustic phonons is not limited to the class of clathrates and related cage-like
materials. For the recently discovered family of Cu$_{\rm 12-x}$M$_{\rm
x}$Sb$_4$S$_{13}$ (M = transition metal) tetrahedrites, optical phonon branches
involving out-of-plane vibrations of the three-fold coordinated Cu ions were
predicted by {\em ab initio} calculations and were suggested as origin of the
low thermal conductivities\cite{Lu13.1}. Interestingly, $\kappa_{\rm ph}$ of
this new material fits perfectly into our universal scaling plot
(Fig.\,\ref{fig:universal}). Other candidate materials are, e.g., PbTe
(Ref.\,\onlinecite{Del11.1}), Bi$_2$Te$_3$ (Ref.\,\onlinecite{Man14.1}), 
BiCu(Se,Te)O (Ref.\,\onlinecite{Vaq15.1}), Cu$_3$SbSe$_3$ (Ref.\,\onlinecite{Wuj14.1}), and rattling-induced superconductors
such as the $\beta$-pyrochlore oxides\cite{Nag09.1}, dodecaborides
\cite{Lor05.1}, or VAl$_{10}$ (Ref.\,\onlinecite{Kli12.1}), and possibly even
amorphous and glassy materials\cite{Shi08.2}. Detailed investigations, such as
presented here, are needed to test whether also in these materials the phonon
Kondo effect is at work.

Phonon Kondo systems transfer heat largely via low frequency phonons of long
mean free paths. As such they are promising {\em intrinsic} ``thermocrystals'',
for applications such as heat waveguides or thermal diodes in the emerging field
of phononics\cite{Mal13.1}.\\[-0.4cm]

\noindent{\bf MATERIALS AND METHODS}

\noindent{\bf Synthesis and structural characterization} \hspace{0.1cm} As
starting material for the single crystal growth of BCGG$x$, two
cylindrical rods with the same nominal composition Ba$_8$Cu$_{4.8}$Ge$_{\rm
41.2-x-y}\square_{\rm y}$Ga$_{\rm x}$ (BCGG$x$) were prepared for each
sample with $x=0.0$, 0.2, 0.5, and 1.0 from high-purity elements using a
high-frequency induction furnace. One rod with 7\,mm in diameter and 60\,mm in
length served as the feed rod, the other one with the same diameter and 20\,mm
in length as the seed for the crystal growth. The single crystal growth was
performed in a 4-mirror furnace equipped with 1000\,W halogen lamps. The pulling
speed of the rod was 3-5\,mm/h. Both rods rotated in opposite direction
(speed:~$\sim$8\,revolutions\,per\,minute) to ensure efficient mixing of the
liquid and a uniform temperature distribution in the molten zone. A pressure of
1.5\,bar Ar was used during the crystal growth.

X-ray powder diffraction data on BCGG$x$ were collected using a
HUBER-Guinier image plate system (Cu K$_{\alpha_1}$, $8^\circ \leq 2\theta \leq
100^\circ$). The lattice parameters (Fig.\,\ref{fig:scattering}a) were obtained
from least squares fits to indexed $2\theta$ values employing Ge ($a_{\rm
{Ge}}=0.5657906$\,nm) as internal standard. \vspace*{0.4cm}

\noindent{\bf Thermal conductivity} \hspace{0.1cm} Commonly used laser flash
methods measure the thermal diffusivity and thus need to be combined with
specific heat and density measurements to calculate the thermal conductivity,
which typically reduces the accuracy of this technique. By contrast, the
3$\omega$ method is an ac technique for direct thermal conductivity
measurements. During a 3$\omega$ experiment the sample is heated locally and
thus, in contrast to steady-state heat-flow experiments, errors due to radiation
at elevated temperatures are reduced to a negligibly low
level\cite{Cah87.1}. Furthermore, this method is insensitive to
geometrical errors. This is because the only geometrical parameter entering is
the length of a line heater. As it is usually prepared by means of photo or
electron-beam lithography and sputtering, its length is very well defined (see below). In fact, the error of our 3$\omega$ thermal conductivity data, which we estimate to be below 5\%, is dominated by uncertainties in the heater resistance and its temperature dependence.

For our studies, the narrow metal line serving as both the heater and the
thermometer had a width of 20\,$\mu$m and a length of 1\,mm, with an uncertainly
of 1\,$\mu$m. To avoid electrical contact between heater and sample a thin layer
of SiO$_2$ was first deposited on the polished sample surfaces by chemical vapor
deposition. Then, a 4\,nm thick titanium sticking layer and the 64\,nm thick 
gold film were sputtered in an Ardenne LS 320 S sputter system. The heater
structures were made by standard optical lithography techniques using a Karl
Suess MJB4 mask aligner.

The metal line was heated by an oscillating current at a circular frequency
$\omega$, which thus leads to a 2$\omega$ temperature oscillation of both the
heater and the sample. Due to the linear temperature dependence of the metallic
heater, the 2$\omega$ temperature oscillation translates into a 3$\omega$
voltage oscillation, which is detected using a lock-in amplifier (7265, Signal
Recovery). Applying the 3$\omega$ method\cite{Cah87.1} to bulk geometry, the
measured in-phase temperature oscillation of the heater/thermometer line is
expected to be linear in logarithmic frequency $f$ as long as the thermal
penetration depth is large compared to the heater half width $b$ and at least
five times smaller than the sample thickness $t$ (see boundaries indicated in
Fig.\,\ref{fig:kappa}a). The thermal conductivity $\kappa$ of the material can
then be extracted from the slope of the in-phase temperature oscillation $\Delta
T$ vs $\log f$.

Prior to the thermal 3$\omega$ voltage detection, the first harmonic and all
related higher harmonics are subtracted from the signal using a
carefully gain and phase calibrated active filter\cite{Cah87.1} based on a
technique that allows to adjust the magnitude and phase of a reference signal as
a function of frequency. In this way the main error sources of a 3$\omega$
experiment, spurious 3$\omega$ signals arising from harmonic distortions of the
involved amplifiers, can be largely eliminated. Using an ultra-precise
100\,$\Omega$ Z-foil resistor (temperature coefficient $\pm$0.05\,ppm/$^\circ$C,
tolerance $\pm$50\,ppm; Vishay) as a sample we found the background signal to be
negligibly small within the entire frequency range (0.5\,Hz to 50\,kHz) of the
experiment. With the same resistor, the accuracy of our voltage controlled AC
constant current source was found to be within $\pm$0.1\,\% within the whole
frequency range and for all tested excitations (100\,$\mu$A to 10\,mA). By
further measurements on different resistors, load dependencies were ruled out.

3$\omega$ measurements were done in the temperature range between 80 and 330\,K.
With this setup, we reproduced the thermal conductivity data on single
crystalline Ba$_8$Ga$_{16}$Ge$_{30}$ of Sales et al.\cite{Sal01.1} within the
error bar, estimated to only 3\% in that work. This precision, which is
remarkable for a steady-state experiment, was reached with special radiation
shields and specific sample geometries\cite{Sal01.1,Sal97.1}.

Below 100\,K the data were completed by additional steady-state heat-flow
experiments. The phonon thermal conductivities of all investigated materials
were calculated by subtracting an electronic contribution, determined using the
Wiedemann-Franz law with a constant Lorenz number of
$L_0=2.44\cdot10^{-8}$\,W$\Omega$K$^{-2}$ and electrical resistivity data
measured on the same samples.\vspace*{0.4cm}

\noindent{\bf Specific heat} \hspace{0.1cm} The specific heat was measured with
a relaxation-type method using the $^4$He specific heat option of a Physical
Property Measurement System (PPMS) from Quantum Design. The addenda was
measured separately prior to each sample measurement.

To study the phonon contribution to $C_\text{p}$, first the electronic
contribution was determined. At low temperatures, the specific heat can be
approximated by $C_\text{p}$/$T$=$\gamma$+$\beta T^2$, where the Sommerfeld
coefficient $\gamma$ represents the electron contribution and the $\beta$
parameter quantifies the Debye-like phonon contribution. For BCGG$x$, the
data below 3.5\,K are very well described by such linear fits (not shown).

Rattling modes, originating from localized oscillations of the guest atoms, can
be revealed by analyzing $C_\text{p}$/$T^3$ vs $\log T$. Within such a
representation, rattling modes appear as bell-shaped contributions on top of a
Debye-like phonon background (not shown). \vspace*{0.4cm}

\noindent{\bf Hall effect and electrical resistivity} \hspace{0.1cm} The
electrical resistivity and the Hall coefficient were determined by a standard
6-wire technique using the horizontal rotator option of a Physical Property
Measurement System (PPMS) from Quantum Design. Temperature dependent Hall effect
measurements were performed in a magnetic field of 9\,T. The Hall resistivity
was confirmed to be linear in fields up to 9\,T at all temperatures down to
2.5\,K. The Hall coefficient was analyzed within a simple one-band model,
$R_{\rm H}=1/ne$. \vspace*{0.4cm}

\noindent{\bf Thermal expansion} \hspace{0.1cm} Measurements of the coefficient
of thermal expansion $\alpha_{\rm{L}}(T) = l^{-1}dl/dT$ were carried out by
using a high-resolution capacitive dilatometer\cite{Pot83.1}, which enables
the detection of length changes $\Delta l \ge 10^{-2}$\,\AA. Relative length
changes were measured along a principle axis of cubic Ba$_8$Ga$_{16}$Ge$_{30}$.
$\alpha_{\rm{L}}(T)$ is obtained by numerical differentiation of the $\Delta
l(T)/l$ data with respect to temperature (Supplementary Material S7).
\vspace*{0.4cm}

\noindent{\bf {\em Ab initio} calculations} \hspace{0.1cm} {\em Ab initio}
density functional theory (DFT) simulations were conducted using the Vienna Ab
initio Simulation Package (VASP)\cite{Kre96.2}, applying the projector augmented
wave method\cite{Kre99.1} and the generalized gradient approximation (GGA) as
proposed by Perdew and Wang (PW91)\cite{Per92.1}. A fully ordered, cubic
54 atom unit cell of Ba$_8$Ga$_{16}$Ge$_{30}$ with Ba at the Wyckoff sites
$2a$ and $6d$, Ge at $6c$ and $24k$, and Ga at $16i$ was investigated, using a
$5\times 5\times 5$ $k$-point mesh and a plane wave cutoff of 500\,eV. For a
fixed lattice constant of 10.74\,\AA, corresponding to the experimental value,
the atomic positions were relaxed. After reaching convergence (residual forces
of less than $10^{-3}$\,eV/\AA), symmetry non-equivalent displacements were
introduced into the relaxed structure. For displacements of 0.02\,\AA, the
restoring forces were determined, again by using the VASP code with the above
described settings. From the obtained forces, the dynamical matrix was extracted
and the alamode\cite{Tad14.1,Tad15.1} code was used to determine the phonon
DOS and the specific heat (Supplementary Fig.\,1, main parts and
insets) in the harmonic approximation.

The thermal expansion coefficient was determined within the quasi-harmonic
approximation. For this purpose, the ground state structure and energy were
determined for a series of unit cells with fixed cubic lattice parameter,
corresponding to both decreased and increased cell volumes. For each of these
volumes the dynamical matrix was then obtained in the same way as described
above for the experimental volume. By using the phonopy code\cite{Tog15.1}
the dynamical matrix could be used to determine the free energy at the given
volume and as a function of temperature. The free energy as a function of the
volume for a given temperature was then fitted to the Vinet equation of state,
again using the phonopy code. Finally, from the free energy minima at a given
temperature, the corresponding equilibrium volume at this temperature can be
extracted, which then allows to access the thermal expansion coefficient. GGA
generally slightly overestimates bonding. Thus, the {\em ab initio} thermal
expansion curves are rescaled to match the high-temperature data (Supplementary
Fig.\,2).

For the thermal conductivity, {\em ab initio} results obtained by Tadano et
al.\cite{Tad15.1} using anharmonic interatomic force constants are used. We
reproduced these calculations to within a factor of 1.1 (the difference is
likely due to a slightly different lattice constant used in our
calculations).\\[-0.4cm]

\noindent{\bf ACKNOWLEDGEMENTS}

We thank J.\,Schalko for preparing insulating SiO$_2$ layers and S.\ Pailh\`es,
V.\ M.\ Giordano, M.\ de Boissieu, E.\ Andrei, and M.\ Foster for insightful
discussions. Financial support from the Austrian Science Fund (FWF projects
TRP176-N22, I623-N16, and DK W1243), the Deutsche Forschungsgemeinschaft (DFG
SPP1386 and SFB/TR49), and the European Research Council (ERC Adv.\ Grant 227378
QuantumPuzzle) is gratefully acknowledged.\\[-0.4cm]


\noindent{\bf AUTHOR CONTRIBUTIONS}

M.I. and S.P. designed the research. X.Y., P.T., and A.P. synthesized the
materials. M.I. and R.S. designed the 3$\omega$ setup. M.I., L.P., X.Y., P.T.,
G.L., S.H., and E.G. performed the measurements. M.L. supervised the thermal
expansion work. M.I. analyzed the data. H.E. performed the {\em ab initio}
calculations. M.I. and S.P. interpreted the results and prepared the manuscript.
All authors contributed to the discussion.

\noindent$^{\ast}$To whom correspondence should be addressed. E-mail:
paschen@ifp.tuwien.ac.at\\[-0.4cm]

\noindent{\bf ADDITIONAL INFORMATION}

Supplementary Information accompanies the paper on XXX (doi: XXX).\\[-0.4cm]

\noindent{\bf COMPETING INTERESTS}

The authors declare no competing interests.
\newpage

\noindent{\bf REFERENCES}


\newpage
\begin{figure}[ht!]
{\large\bf \hspace{-8.3cm} a \hspace{8cm} b}
\vspace{-0.8cm}

\centering	
\subfigure{\includegraphics[width=0.42\textwidth]{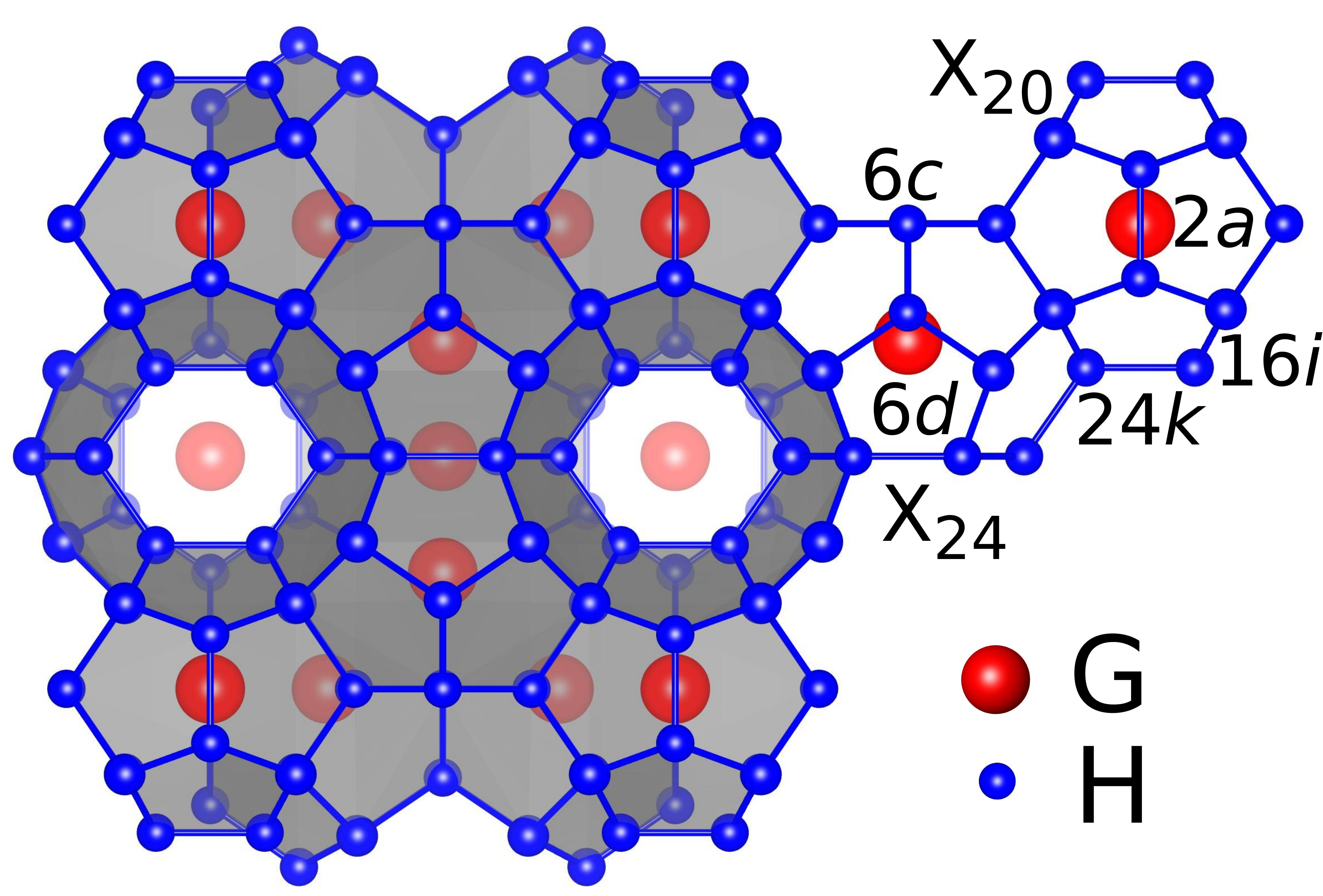}}\hfill
\subfigure{\includegraphics[width=0.40\textwidth]{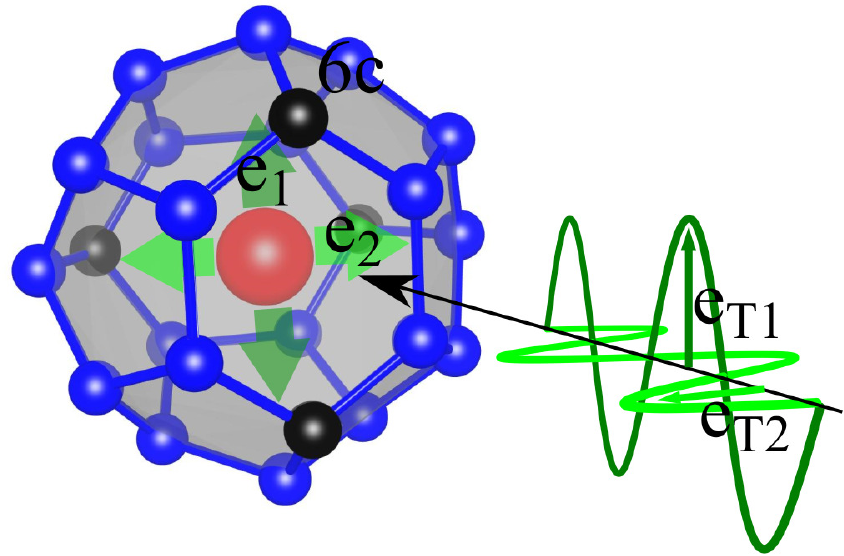}}	\\

{\large\bf \hspace{-9.8cm} c \hspace{5.5cm} d}

\subfigure{\vspace{-1cm}\includegraphics[width=0.35\textwidth]{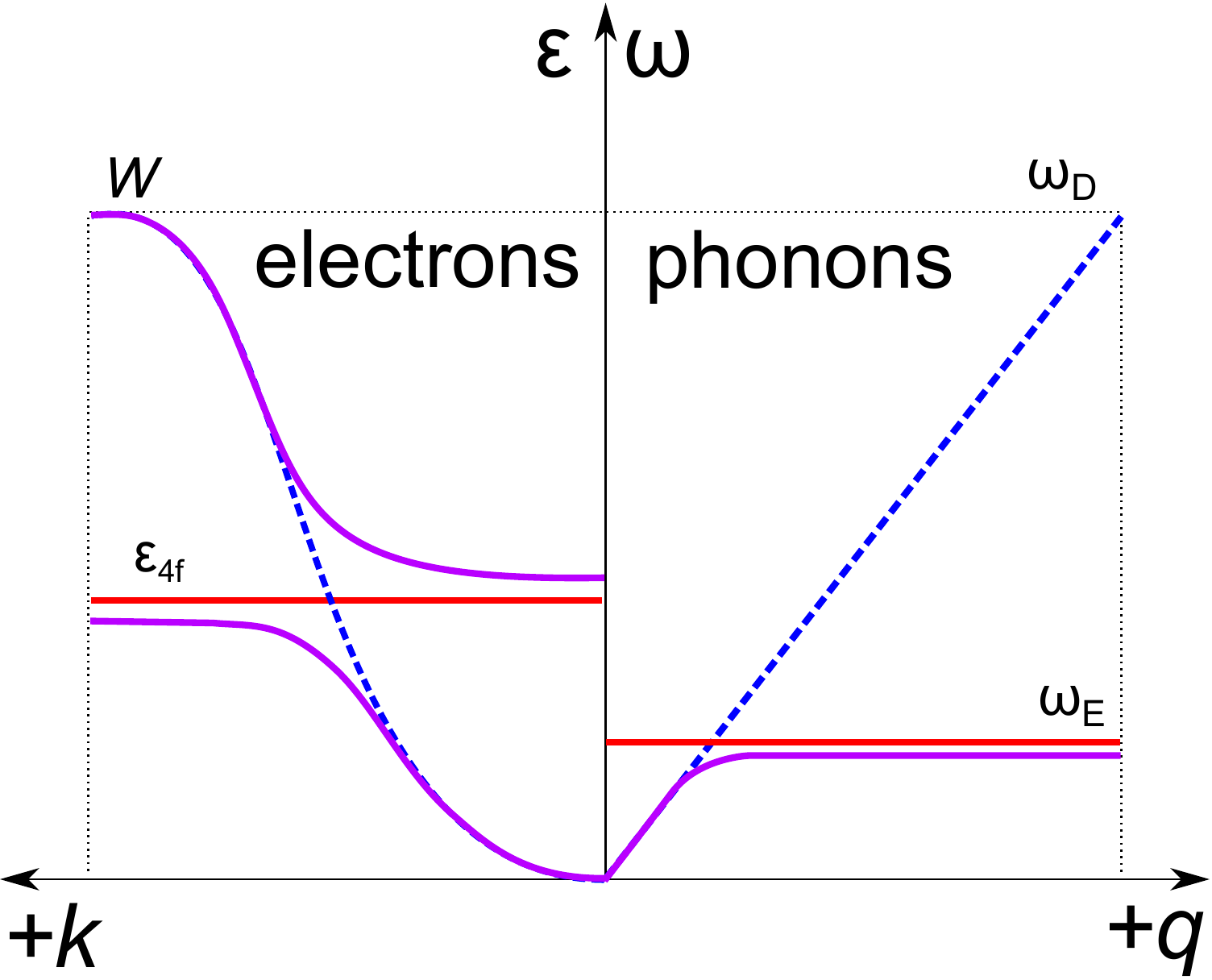}}\hfill
\subfigure{\includegraphics[width=0.6\textwidth]{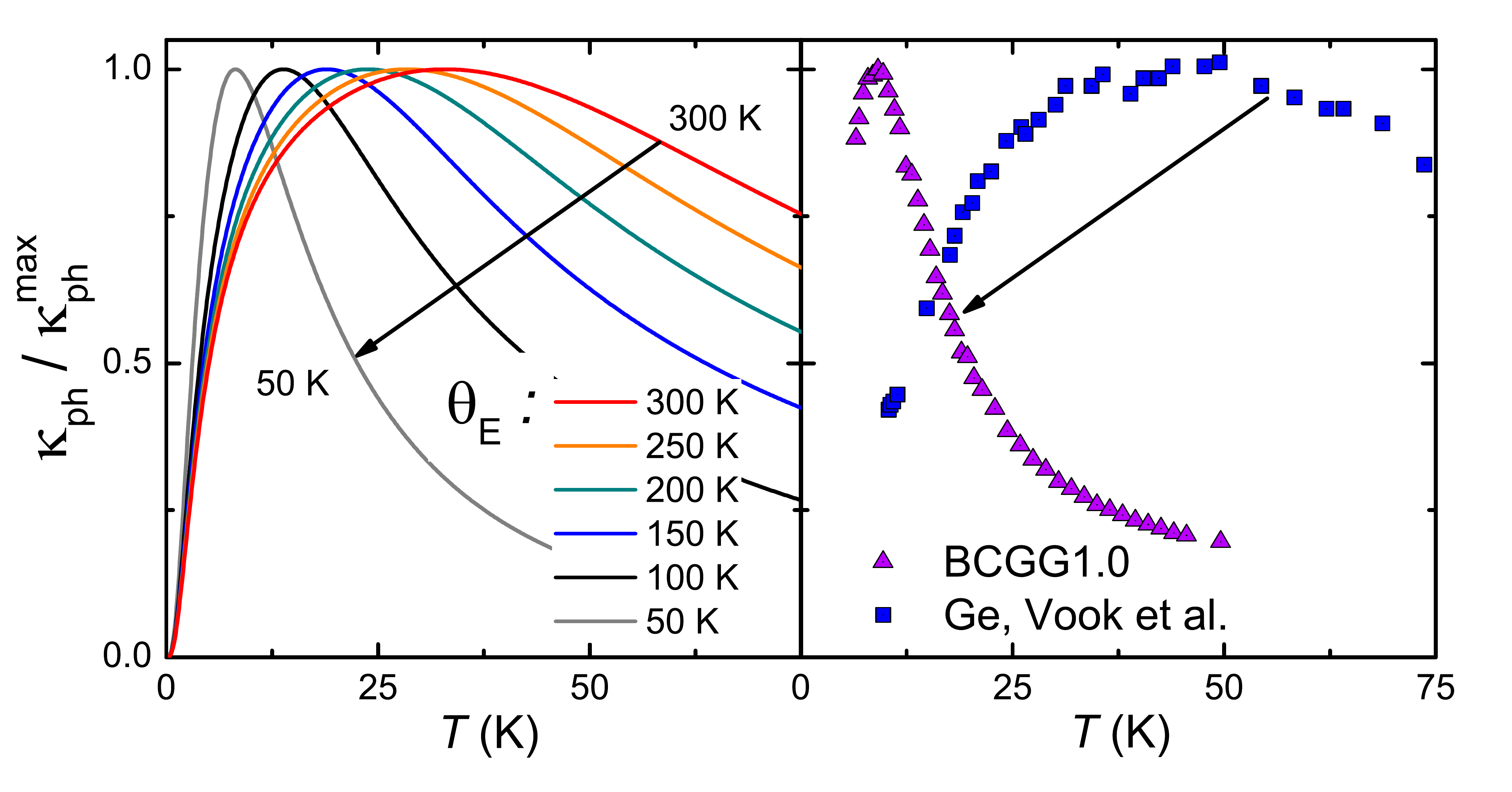}}\\

{\large\bf \hspace{-7.3cm} e \hspace{8cm} f}

\subfigure{\includegraphics[height=0.29\textheight]{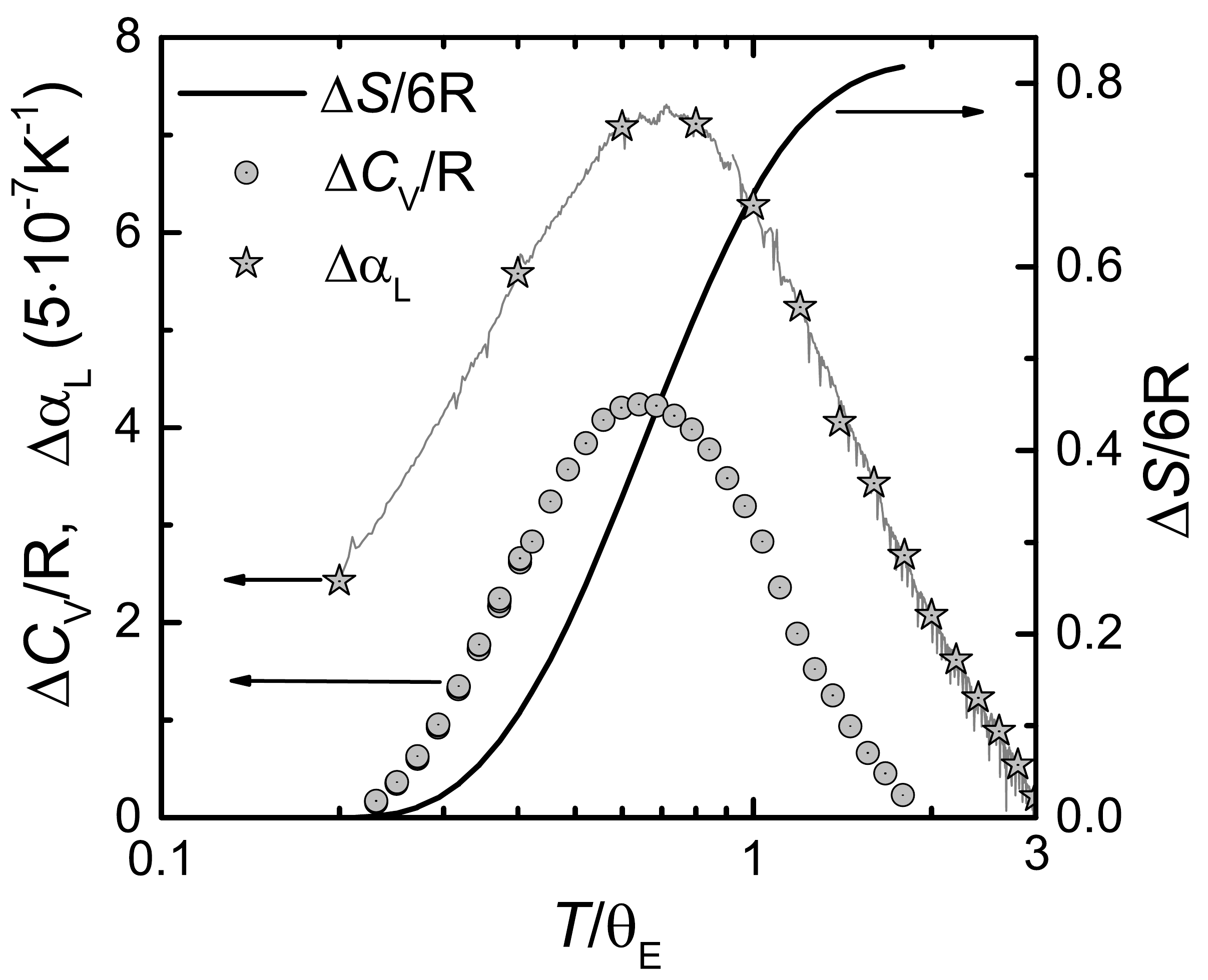}}\hfill
\subfigure{\includegraphics[height=0.29\textheight]{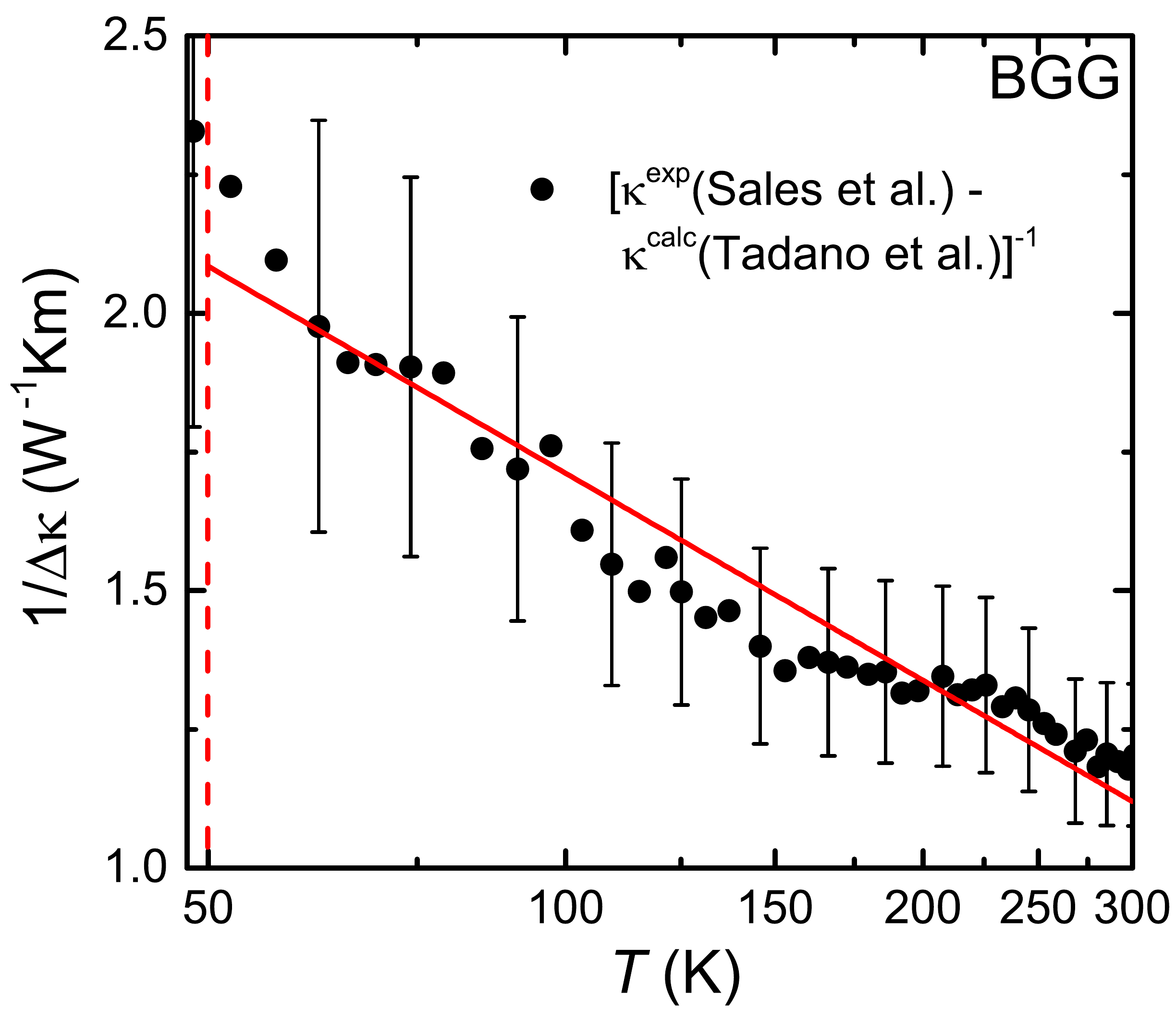}}\\

\newpage

\end{figure}

\begin{figure}[t!]
\caption{{\bf Comparison of spin and phonon Kondo effect.} ({\bf a}) Type-I
clathrate crystal structure G$_8$H$_{46}$. Per unit cell 8 guest atoms (G, red)
are situated in two different cages (smaller dodecahedra X$_{20}$ and larger
tetrakaidecahedra X$_{24}$) built by the host (H, blue) atoms. The different
crystallographic sites are labeled. ({\bf b}) Sketch of an X$_{24}$ cage with
the two soft rattling directions $e_1$ and $e_2$ within the easy plane parallel
to the two hexagonal faces of the cage. Each soft direction is parallel to the
longest secant of one of the two hexagons, both defined by the $6c$ site (black
atoms) of the structure. Incoming acoustic phonons with the two different
transverse polarizations $e_{\rm{T1}}$ and $e_{\rm{T1}}$ are sketched by the
sinusoidal waves. ({\bf c}) Schematic dispersion relations for electrons in
heavy electron systems (left) and phonons in an analogously defined ``heavy''
phonon system (right). \SBP{The blue and red curves represent the non-interacting dispersive and localized entities, respectively, the violet curves the hybridized interacting states. As function of temperature, the system evolves from non-interacting well above the Kondo temperature to interacting well below it.} For simplicity, phonon branches above $\omega_{\rm E}$
are neglected. This is justified in real clathrates by the presence of multiple
Einstein temperatures, resulting in multiple
anticrossings\cite{Euc12.1,Pai14.1}. The Debye model (blue line, right) assumes
$\omega = v_{\rm s} q$ where $v_{\rm s}$ is the sound velocity. The new
dispersion relation (violet, right) is characterized by the group velocity
$v_{\rm g} = \partial\omega/\partial q$. It equals the sound velocity only at
low wave vectors and frequencies. ({\bf d}) Temperature dependent phonon thermal
conductivity, normalized to its maximum, calculated using a modified Callaway
model (Supplementary Information S1) for various Einstein temperatures
$\Theta_{\rm E}$ (left) and corresponding data for the Ge-based type-I clathrate
BCGG1.0 (Supplementary Table~1) and elemental Ge, electron irradiated and
annealed at 77\,K to similar defect densities\cite{Voo65.1} (right). ({\bf e})
Specific heat and thermal expansion phonon Kondo anomaly, obtained by
subtracting the experimental from the theoretical specific heat and thermal
expansion curves of Ba$_8$Ga$_{16}$Ge$_{30}$ (left axis, see Supplementary
Figs.\,1b and 2b and Materials and Methods) and the corresponding entropy (right
axis). ({\bf f}) The inverse difference of calculated\cite{Tad15.1} and
experimental\cite{Sal01.1} phonon thermal conductivities of
Ba$_8$Ga$_{16}$Ge$_{30}$ (Supplementary Fig.\,3b), showing the $-\ln T$ hallmark
(full red line) of incoherent Kondo scattering in the spin Kondo effect
above the Kondo temperature (dashed vertical line). The error bars 
represent the error of $\pm 3$\% specified for the thermal conductivity
data\cite{Sal01.1}.}
\label{fig:kappa2}
\end{figure}

\newpage

\newpage

\begin{figure*}[t!]
{\large\bf \hspace{-5cm} a \hspace{4.5cm} b \hspace{4.5cm} c}
\vspace{-0.1cm}

\centering
\subfigure{\includegraphics[width=0.33\textwidth]{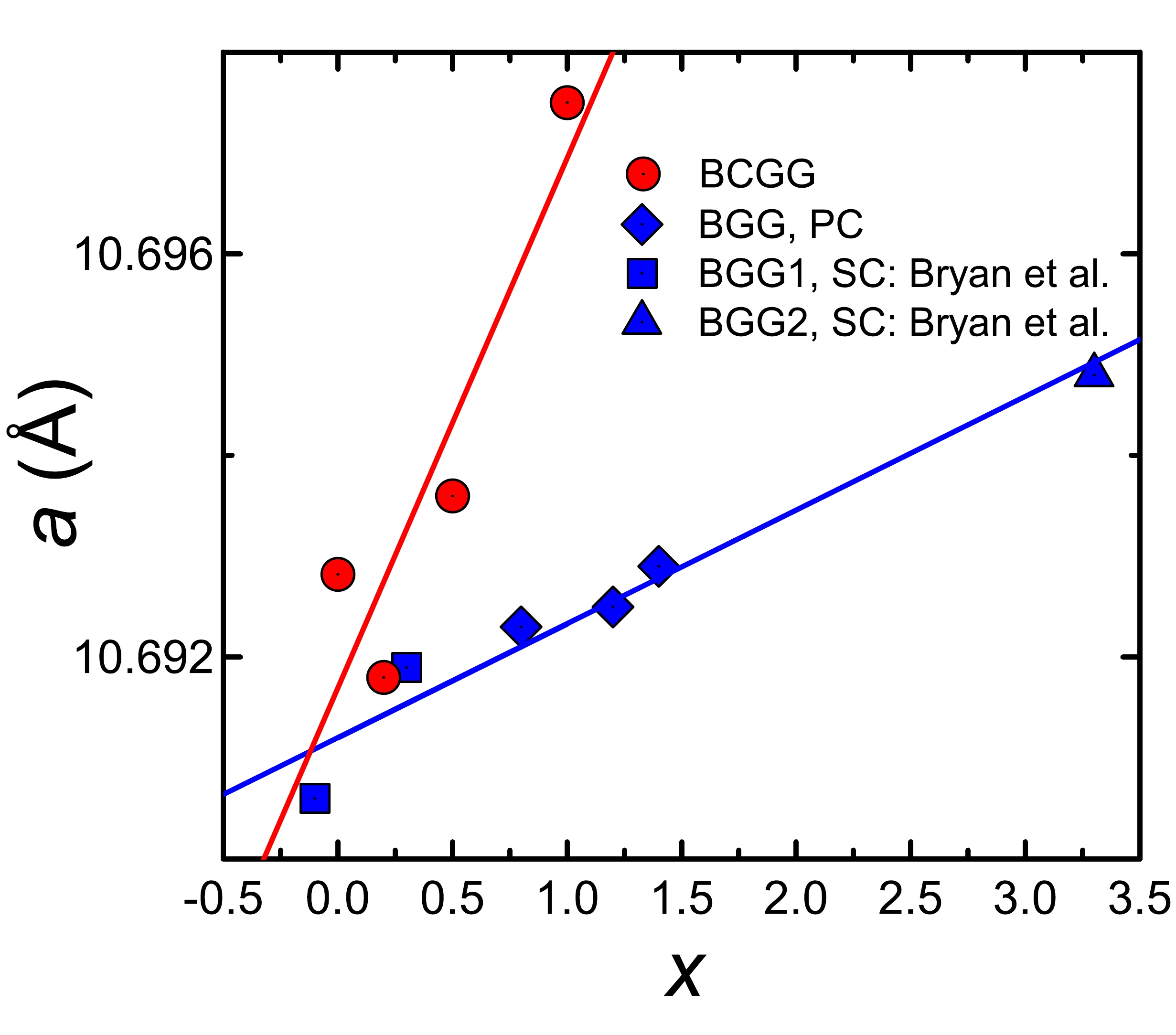}}\hfill
\subfigure{\includegraphics[width=0.33\textwidth]{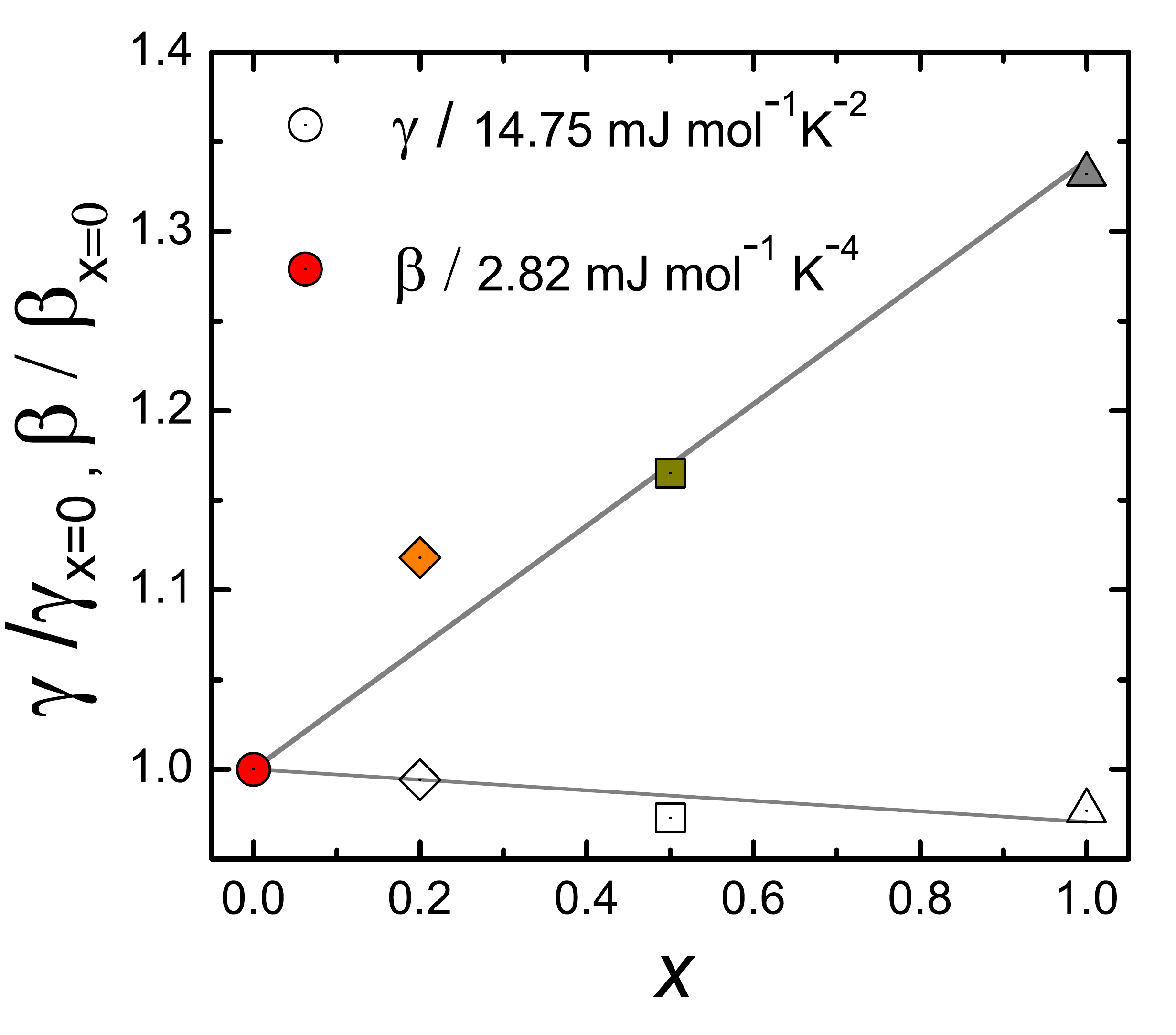}}\hfill
\subfigure{\includegraphics[width=0.33\textwidth]{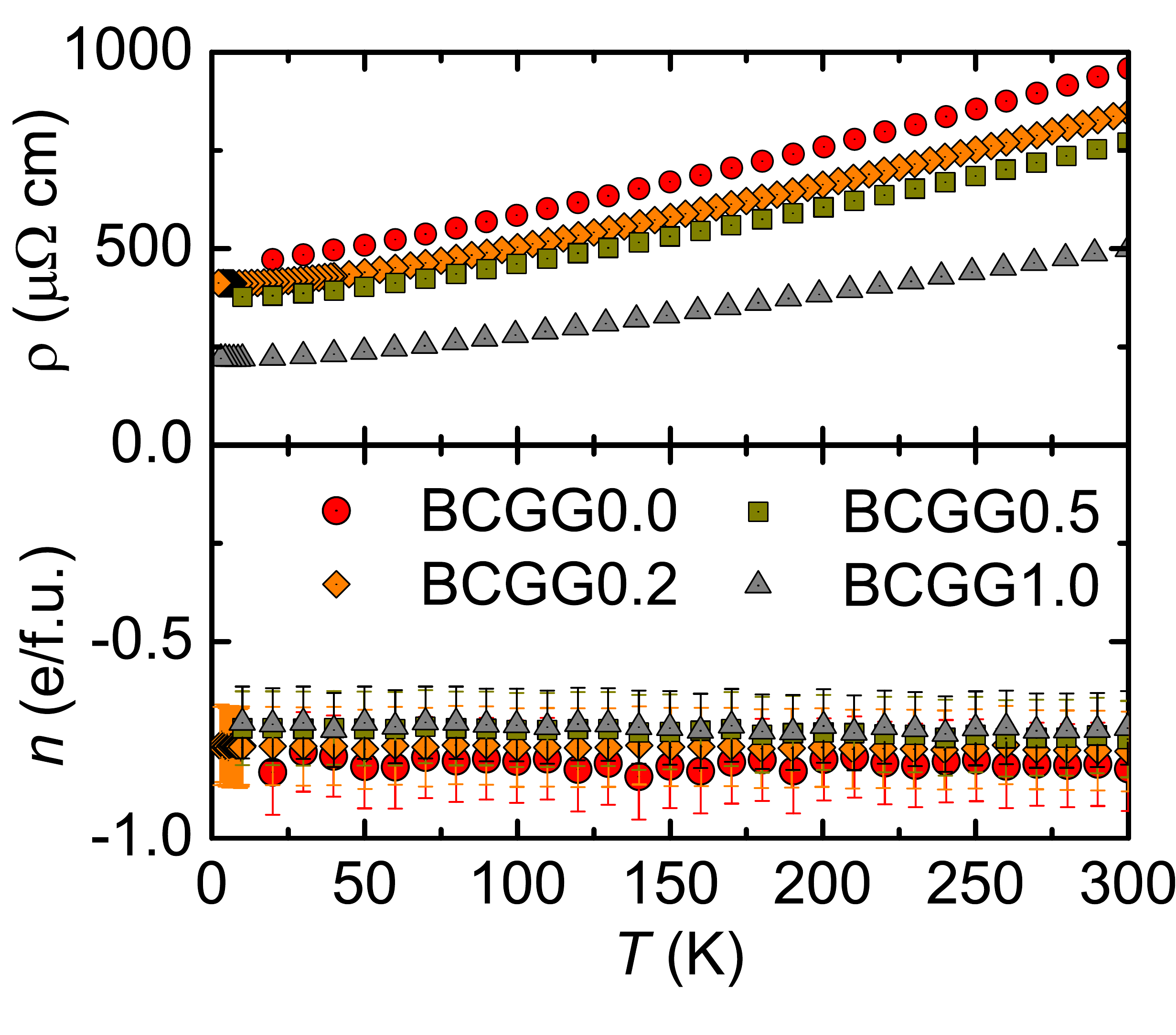}}\\
\subfigure{\includegraphics[width=0.627\textwidth]{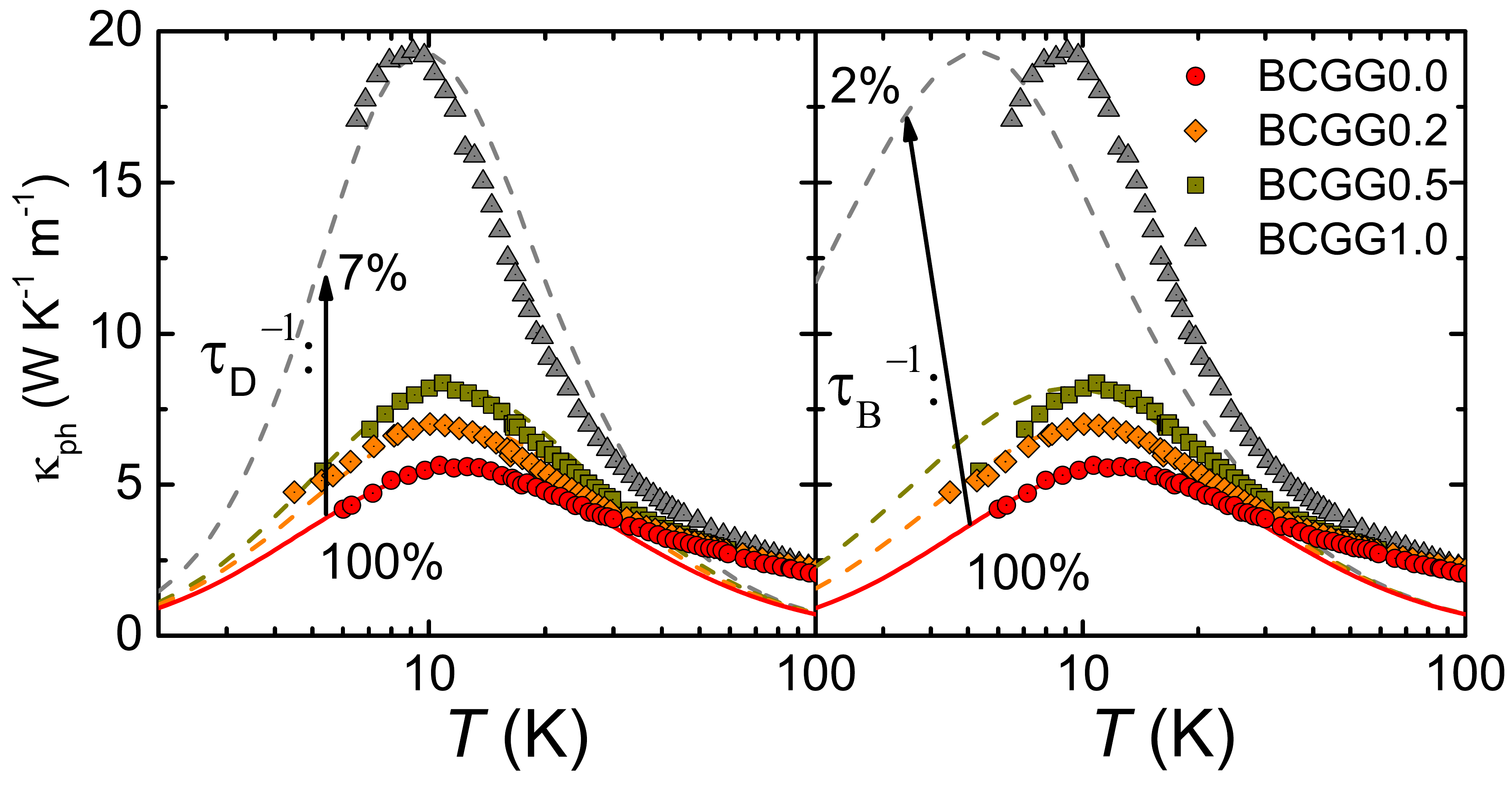}}\hfill 
\subfigure{\includegraphics[width=0.373\textwidth]{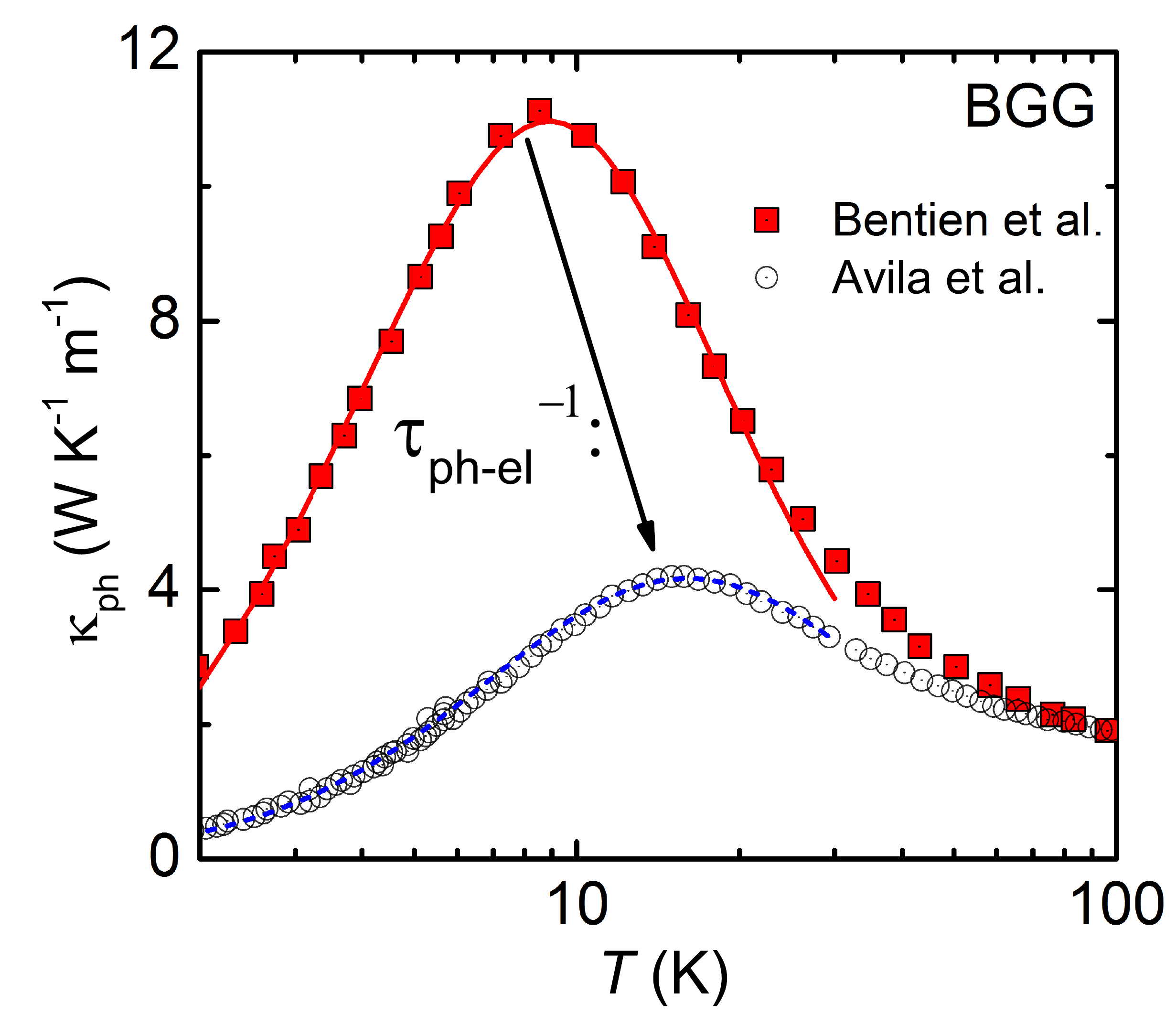}}
\vspace{-7.3cm}

{\large\bf \hspace{-5.8cm} d \hspace{9.2cm} e}
\vspace{5.5cm}

\caption{{\bf Scattering contributions in clathrate single crystal series.}
({\bf a}) Lattice parameter vs $x$ for Ba$_8$Cu$_{4.8}$Ge$_{\rm
41.2-x-y}\square_{\rm y}$Ga$_{\rm x}$ (BCGG$x$, Supplementary Table~1) and
comparison with the series Ba$_8$Ga$_{\rm 14+x}$Ge$_{\rm 32-x}$ (BGG are
polycrystals synthesized by us, BGG1 are single crystals 4 and 6 from
Ref.\,\onlinecite{Bry02.1}, and BGG2 is a single crystal from
Ref.\,\onlinecite{Bry99.1}), which was shifted downwards by 89.6\,m\AA\ for
better readability. For the linear fits see Supplementary Information S3. ({\bf
b}) Coefficients $\gamma$ and $\beta$ of linear fits $C_{\rm p}/T = \gamma +
\beta T^2$ to the specific heat data, normalized to their respective value for
$x=0$, of Ba$_8$Cu$_{4.8}$Ge$_{\rm 41.2-x-y}\square_{\rm y}$Ga$_{\rm x}$ below
3.5\,K (not shown) vs $x$. ({\bf c}) Electrical resistivity (top) and charge
carrier concentration (bottom), determined from the Hall coefficient in a
one-band model, $R_{\rm H} = 1/ne$, as function of temperature for all
Ba$_8$Cu$_{4.8}$Ge$_{\rm 41.2-x-y}\square_{\rm y}$Ga$_{\rm x}$ samples. ({\bf
d}) Phonon thermal conductivity vs temperature for all Ba$_8$Cu$_{4.8}$Ge$_{\rm
41.2-x-y}\square_{\rm y}$Ga$_{\rm x}$ samples (symbols), together with fits with
the modified Callaway model (Supplementary Information S1) to the $x=0$ data
(red full line) and simulations (dashed lines) using the parameters of the $x=0$
fit except for the defect scattering rate $\tau_{\rm D}^{-1}$ (left) and the boundary scattering rate $\tau_{\rm B}^{-1}$ (right). The
respective scattering rate was decreased to the indicated percentage in
direction of the arrow. The speed of sound of each sample was determined from
$\beta$, the Einstein temperature $\Theta_{\rm E}$ from bell shaped
contributions to $C_{\text{p}}/T^3$ versus log$T$. ({\bf e}) Phonon thermal
conductivity vs temperature for two different} 
\label{fig:scattering}
\end{figure*}

\addtocounter{figure}{-1}
\begin{figure*}[t!]
\caption{(cont.) Ba$_8$Ga$_{\rm 16-x}$Ge$_{\rm
30+x}$ samples from the literature\cite{Ben04.1,Avi06.1}, with
lines corresponding to our fit to the data from Ref.\,\onlinecite{Ben04.1} and a
simulation using essentially the same parameters except for a strongly enhanced
phonon-electron scattering rate $\tau^{-1}_{\rm
ph-el}$ for the data from Ref.\,\onlinecite{Avi06.1}.} 
\end{figure*}

\newpage

\begin{figure*}[ht!]
\centering
\subfigure{\includegraphics[width=0.85\textwidth]{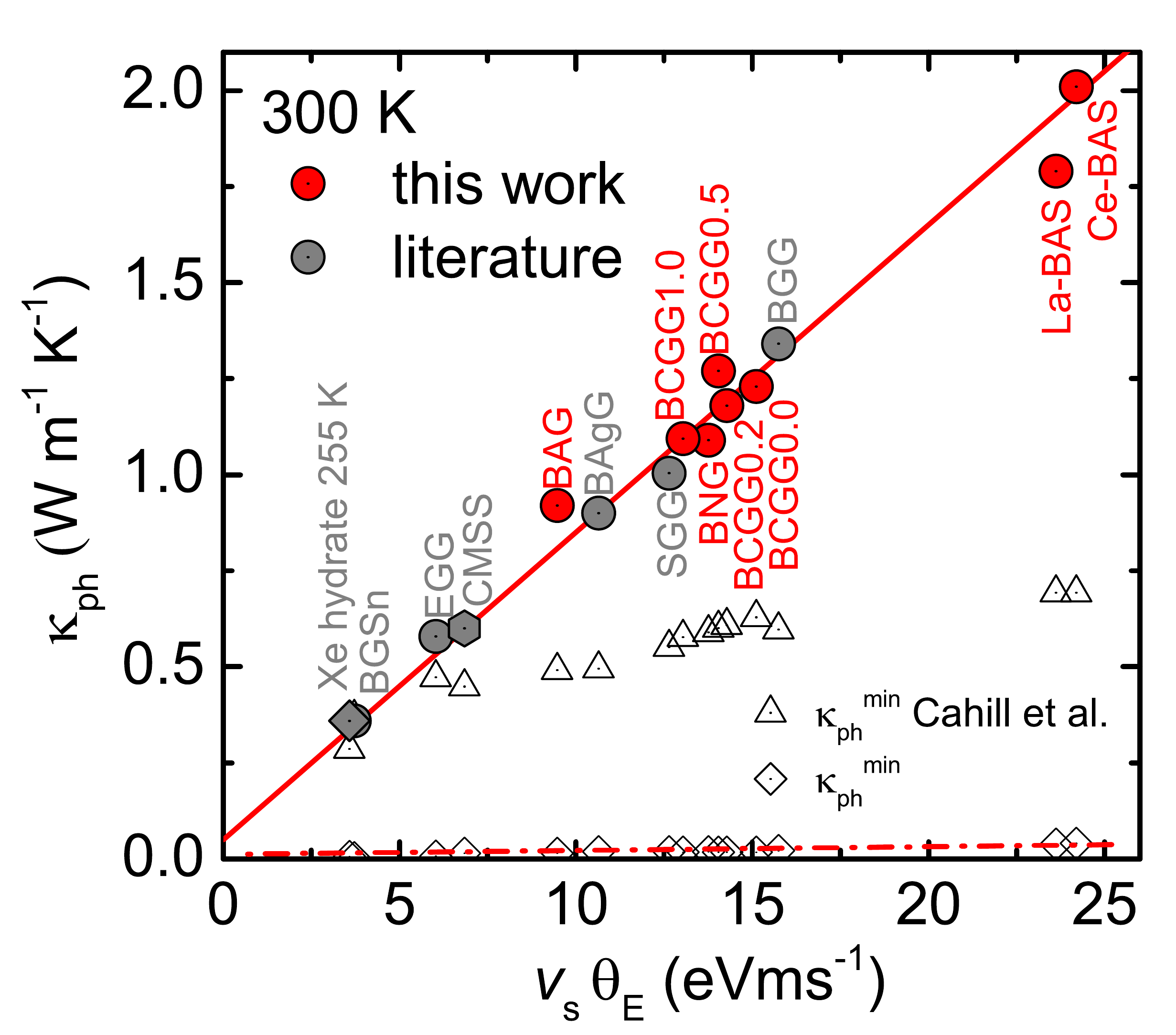}}
\caption{{\bf Universal scaling of the intrinsic phonon thermal conductivity of
phonon Kondo compounds.} Phonon thermal conductivities vs the product of
sound velocity and lowest-lying Einstein temperature (Supplementary
Information S5) for single crystalline intermetallic clathrates measured within
this work (full circles), together with published data on intermetallic
clathrates (full squares), a Xe-filled ice hydrate, and the tetrahedrite
Cu$_{10.6}$Mn$_{1.4}$Sb$_4$S$_{13}$ (CMSS, see Supplementary Table~1 for all
sample compositions and references). The open triangles are the minimum
$\kappa_{\rm ph}$ values derived using $\kappa_{\rm min}$ of
Ref.\,\onlinecite{Cah92.1}. The open squares are the minimum $\kappa_{\rm ph}$
values estimated using Supplementary Eqn.\,9.} 
\label{fig:universal}
\end{figure*}

\end{document}


\begin{center}
\vspace{1cm}

{\Large Supplementary Material for}
\vspace{0.5cm}

{\large\bf{Kondo-like phonon scattering in thermoelectric clathrates}}
\vspace*{0.8cm}

M.~Ikeda, H.~Euchner, X.~Yan, P.~Tome\v{s}, A.~Prokofiev, L.~Prochaska, G.~Lientschnig, R.~Svagera, S.~Hartmann, E.~Gati, M.~Lang, and S.~Paschen

\end{center}
\vspace*{0.8cm}

\noindent\hspace{1cm} {\bf Supplementary Discussion S1\,-\,S8}

\noindent\hspace{1cm} {\bf Supplementary Equations S1\,-\,S11}

\noindent\hspace{1cm} {\bf Supplementary Table S1}

\noindent\hspace{1cm} {\bf Supplementary Figures S1\,-\,S4}

\noindent\hspace{1cm} {\bf Additional References}

\newpage
\noindent{\bf\large Supplementary Discussion}\\
\vspace{-0.4cm}

\noindent{\bf S1. Modified Callaway model}

\noindent The Callaway model\cite{Cal59.1} is a phenomenological treatment of the phonon thermal conductivity of solids. It is derived from the Boltzmann equation for an isotropic crystal with dispersionless acoustic phonon branches, assuming that all phonon scattering processes can be represented by frequency-dependent relaxation times. Both resistive three-phonon Umklapp scattering processes and total momentum conserving normal processes are considered. Its simplicity and early success in describing simple materials such as Germanium made it a popular choice to model experimental thermal conductivity data, even for more complex materials.

To model the low-temperature thermal conductivity of type-I clathrates we propose here a modified form of the Callaway model, namely
\begin{equation}\tag{S1}
\begin{split}
\kappa_{\text{ph}}  = & \frac{k_\text{B}^4 \cdot T^3}{2 \cdot \pi^2 \cdot v_\text{s} \cdot \hbar^3} \cdot \left[\int^{\frac{\Theta_\text{E}}{T}}_0 {\tau \cdot \frac{x^4 \cdot e^x}{\left(e^x-1\right)^2} \cdot dx} +\frac{I_1^2}{I_2}\right]\quad,\\
\mbox{with}\quad I_1 & =\int^{\frac{\Theta_\text{E}}{T}}_0 {\frac{\tau}{\tau_\text{N}} \cdot \frac{x^4 \cdot e^x}{\left(e^x-1\right)^2} \cdot dx}\quad, \\
I_2 & =\int^{\frac{\Theta_\text{E}}{T}}_0 {\frac{1}{\tau_\text{N}} \cdot
\left(1-\frac{\tau}{\tau_\text{N}}\right) \cdot \frac{x^4 \cdot
e^x}{\left(e^x-1\right)^2} \cdot dx} \quad , \quad \mbox{and}\\
\tau^{-1}  & =\tau_{\text{N}}^{-1}+\tau^{-1}_\text{D}+\tau^{-1}_\text{B}+\tau^{-1}_\text{U}+\tau^{-1}_{\text{ph-el}}\quad.
\label{eq:callaway}
\end{split}
\end{equation}
It differs from the original Callaway model only by the substitution of
the Debye temperature $\Theta_{\rm D}$ by the characteristic temperature of the
lowest-lying Einstein contribution $\Theta_{\rm E}$. This approximation
effectively cuts off all phonon modes above the lowest-lying optical mode, which
is justified by the significantly smaller group velocities of the higher-lying
modes\cite{Euc12.1,Pai14.1}. At temperatures below $\Theta_{\rm E}/2$, where
we performed quantitative analyses (Fig.\,1d, Fig.\,3d,\,e of main part), errors
arising from this approximation are negligible. The empirical scaling of the
intrinsic phonon thermal conductivities of a large number of clathrates with
only $v_\text{s}\Theta_{\rm E}$ observed at 300\,K (Fig.\,4 of main part)
further supports this approximation because $\Theta_{\rm E}$ does not contain
any information on the phonon spectrum above $\Theta_{\rm E}$.

Normal processes are usually taken into account by 
\begin{equation}\tag{S2}
\tau_\text{N}^{-1} = N \omega^a T^b \quad .\\
\label{eq:tauN}
\end{equation}
However, for complex crystal structures or materials including a high
defect concentration these processes have been shown to be
negligible\cite{Ase97.1} permitting us to omit the corresponding 
$I_1^2/I_2$ term in Eqn.\,\ref{eq:callaway}, in our analyses. At low
temperatures, we describe defect scattering by Rayleigh scattering\cite{Kle56.1}
\begin{equation}\tag{S3}
\tau^{-1}_\text{D} = D \cdot \omega^4
\label{eq:BDa}
\end{equation}
and boundary scattering by the frequency independent scattering
rate\cite{Kle56.1}
\begin{equation}\tag{S4}
\tau^{-1}_\text{B} = B \quad .
\label{eq:BDb}
\end{equation}
In analogy with the expression for the scattering rate due to three phonon
Umklapp scattering within the Debye approximation\cite{Kle56.1} 
\begin{equation}\tag{S5}
\tau_\text{U}^{-1}\propto \omega^2 e^{-\frac{\Theta_\text{D}}{2T}} \left[ 1+6\frac{T}{\Theta_\text{D}}+24\left( \frac{T}{\Theta_\text{D}}\right)^2+48\left( \frac{T}{\Theta_\text{D}}\right)^3 \right] \quad \mbox{for} \quad T<\Theta_\text{D}\quad,
\label{eq:tauUklemens}
\end{equation}
we derived\red{\cite{Ike15.1}}
\begin{equation}\tag{S6}
\tau_\text{U}^{-1} \propto \frac{\omega^2}{\alpha^2} e^{-\frac{\Theta_\text{E}}{\alpha T}} \left[1+ 2 \frac{\alpha T}{\Theta_\text{E}} +2 \left(\frac{\alpha T}{\Theta_\text{E}}\right)^2\right] \quad \mbox{for} \quad T<\Theta_\text{E}\quad
\label{eq:tauUnew}
\end{equation}
for acoustic phonons scattering off a low-lying flat optical phonon mode with
energy $k_\text{B} \Theta_\text{E}$. The exponential term of
$\tau_\text{U}^{-1}$ derived within the Debye approximation is scaled by
$\Theta_\text{D}$. By contrast, in the model proposed here, the Einstein
temperature $\Theta_\text{E}$ determines the Umklapp scattering rate. As in the
original form\cite{Kle56.1}, the parameter $\alpha$ scales the energy of the
second phonon to be able to participate in three phonon Umklapp scattering. The
upper limit of $\alpha$ is determined by the dispersion of the acoustic branch.
$\alpha$ values around 2.3 obtained from fitting $\kappa_{\rm ph}$ of the BCGG
series (not shown) are reasonable \cite{Kle51.1,Kle56.1}.

An alternative relation for $\tau_\text{U}^{-1}$ of type-I clathrates and
related materials is obtained\red{\cite{Chr16.1}} by substituting $\Theta_\text{D}$ by
$\Theta_\text{E}$ in the empirical expression\cite{Sla64.1}
\begin{equation}\tag{S7}
\tau_\text{U}^{-1} = U \omega^\alpha e^{-\frac{\Theta_\text{D}}{bT}}
\left(\frac{T}{\Theta_\text{D}}\right)^\beta \quad .
\label{eq:Sla}
\end{equation}
Phonon-electron scattering is taken into account by the expression\cite{Pip55.1}
\begin{equation}\tag{S8}
\tau^{-1}_{\text{ph-el}} = \frac{\pi n m^* v_\text{F}\omega}{6 \rho v_\text{s}} \quad \mbox{with} \quad q < 2k_\text{F} \quad \mbox{and} \quad q
l_\text{e} >> 1 \quad,
\label{eq:pip}
\end{equation}
where $n$ is the charge carrier concentration, $m^*$ is the effective electron
mass, $v_\text{F}$ is the Fermi velocity, $\rho$ is the mass density, and
$l_\text{e}$ is the electron mean free path.

Finally we suggest a redefinition of Cahill's minimum phonon thermal
conductivity\cite{Cah92.1} to the new form
\begin{equation}\tag{S9}
\kappa_{\text{ph}}^{\text{min}} = \frac{k_\text{B}^4 \cdot T^3}{2 \cdot \pi^2 \cdot v_\text{g} \cdot \hbar^3} \cdot \int^{\frac{\Theta_\text{E}}{T}}_0 {\frac{x^3 \cdot e^x}{\left(e^x-1\right)^2} \cdot dx} \quad.
\label{eq:kappaminCahill}
\end{equation}
Cutting off phonon modes above $\Theta_\text{E}$ is justified by the same
arguments as in Eqn.\,\ref{eq:callaway}. A direct confirmation of the need for a
new $\kappa_{\text{ph}}^{\text{min}}$ limit is the experimental observation of
the violation of the original (Cahill) limit (see section Materials above).
\vspace{0.4cm}

\noindent{\bf S2. Materials}

\noindent Figure\,1d right compares the phonon thermal conductivity of the
single crystalline clathrate BCGG1.0 with elemental Ge. Both materials possess a
similar Debye temperature (304\,K for BCGG1.0, 362\,K for Ge,
Ref.\,\onlinecite{Bry61.1}). In both cases, phonon-electron scattering and
boundary scattering are expected to be negligible. In contrast to the complex
structure of BCGG1.0 built up of atoms with severely different masses, single
crystalline Ge typically exhibits few defects and thus a very high thermal
conductivity. Therefore, the phonon thermal conductivity of an
electron-irradiated and post-annealed Ge sample (Ge\,II annealed at 77\,K,
Ref.\,\onlinecite{Bry61.1}) was chosen for comparison.

Figure\,3e compares the phonon thermal conductivity of two type-I clathrate
single crystals with nominal composition Ba$_8$Ga$_{16}$Ge$_{30}$.
Phonon-electron scattering can only occur for phonons with a wave vector $q$
smaller than twice the Fermi wave vector $k_{\rm F}$. Using the charge carrier
concentration $n = 0.84 \cdot 10^{27}$\,e$^-$/m$^3$ and the sound velocity
calculated from the Debye temperature $\Theta_{\rm D} = 324$\,K reported for
crystal II of Ref.\,\onlinecite{Ben04.1} we estimate that phonon-electron
scattering should occur below 140\,K.

For a few type-I clathrates\cite{Yan11.1,Sui15.1} and
skutterudites\cite{Rog13.1,Rog15.1}, phonon thermal conductivities have been
reported that violate the Cahill $\kappa_{\rm{ph}}^{\rm{min}}$ limit, even
though this was not explicitly stated in these works. For nanostructured
Ba$_8$Cu$_x$Si$_{46-x}$ ($x=3.6$, 3.8) prepared by ballmilling and hotpressing
$\kappa_{\rm{ph}}$(800\,K$) = 0.2$\,W/Km was found \cite{Yan11.1}. With the
reported Debye temperature $\Theta_{\rm D} = 420$\,K we calculate a Cahill
$\kappa_{\rm{ph}}^{\rm{min}}$ value of 0.85\,W/Km. A less drastic violation
seems to occur for K$_8$Ga$_8$Si$_{38}$ with $\kappa_{\rm{ph}}$(300\,K$) =
0.5$\,W/Km (Ref.\,\onlinecite{Sui15.1}). In the absence of reported $\Theta_{\rm
D}$ values we estimate a Cahill $\kappa_{\rm{ph}}^{\rm{min}}$ value of 0.7\,W/Km
as lower boundary by using $\Theta_{\rm D} = 360$\,K of the heavier-element
Si-based clathrates La-BAS and Ce-BAS (Table S1) as lower
$\Theta_{\rm D}$ boundary. Likewise, also the high-pressure torsion-treated
skutterudites DD$_x$Fe$_3$CoSb$_{12}$ with $x=0.6$ (Ref.\,\onlinecite{Rog13.1})
and $x=0.68$ (Ref.\,\onlinecite{Rog15.1}) show ultralow phonon thermal
conductivities that appear to violate the respective Cahill
$\kappa_{\rm{ph}}^{\rm{min}}$ limits.

Figure\,S1a shows the specific heat at constant volume ($C_{\rm V}$) of Si
versus temperature. The data were calculated from experimental specific heat
values taken at constant pressure ($C_{\rm p}$, Ref.\,\onlinecite{Des85.1}) and
the temperature dependent volumetric thermal expansion coefficient ($\alpha_{\rm
V}$, Ref.\,\onlinecite{Oka84.1}) using the expression $C_{\rm p} - C_{\rm V}
=\alpha_{\rm V}^2 \cdot T \cdot V_{\rm mol} \cdot B$, where $V_{\rm mol}$ is the
molar volume and $B=97.8$\,GPa is the bulk modulus of Si
(Ref.\,\onlinecite{Hal67.1}). For Ba$_8$Ga$_{16}$Ge$_{30}$ (BGG, Fig.\,S1b), the
thermal expansion coefficient was taken from Ref.\,\onlinecite{Chr06.1} and the
bulk modulus from Ref.\,\onlinecite{Oka08.2}. \vspace{0.4cm}

\noindent{\bf S3. Filling of vacancies by Ga in Ba$_8$Cu$_{4.8}$Ge$_{\rm
41.2-x-y}\square_{\rm y}$Ga$_{\rm x}$}

\noindent Here we show that with increasing Ga content $x$ in
Ba$_8$Cu$_{4.8}$Ge$_{\rm 41.2-x-y}\square_{\rm y}$Ga$_{\rm x}$ the content $y$
of vacancies $\square$ is successively reduced. We assume that, as in related
systems\cite{Ngu10.1,Zha11.1,Zei11.2}, both Cu and the vacancy occupy the 6$c$
site (Fig.\,1a,b). With the multiplicity 6 of this site and 4.8 Cu atoms this
limits the maximum amount of vacancies to 1.2 per unit cell. The linear
dependence of the lattice parameter $a$ on the Ga content $x$ with the slope
$\Delta a/\Delta x = 4.8$\,m\AA\ (Fig.\,3a) suggests that Ga successively fills
the vacancies and that we thus have $y=1.2-x$. In detail, this slope is well
accounted for by a substitution of Ge for $\square$ with $\Delta a/\Delta x =
3.46$\,m\AA\ (Ref.\,\onlinecite{Ngu10.1}) in a first step, and by a substitution
of Ga for Ge with $\Delta a/\Delta x = 1.13$\,m\AA\ (blue line in Fig.\,3a) in a
second step. Both steps together yield a slope of $4.59$\,m\AA\, ($= 3.46 +
1.13$), in good agreement with experiment (red line in Fig.\,3a).

The filling of the vacancies by Ga is corroborated by our low-temperature
specific heat $C_{\rm p}$ and electrical transport (resistivity $\rho$ and Hall
coefficient $R_{\rm H}$) measurements. Whereas the prefactor $\beta$ of the
phonon $T^3$ term of $C_{\rm p}(T)$ increases strongly with the rate
$\Delta\beta/\Delta x \approx 30$\,\% (Fig.\,3b), the electronic Sommerfeld
coefficient $\gamma$ (Fig.\,3b) and the charge carrier concentration $n$
(Fig.\,3c bottom) are essentially independent of $x$. This discards Ga
substituting either Cu or Ge which, according to the Zintl-Klemm
concept\cite{Sch85.1} (see S4 below), would change the charge carrier
concentration by +2 or -1 electron per formula unit, respectively, and suggests
that the formal charge of the vacancy is -1, just as that of Ga. The observed
increase of the charge carrier mobility $\mu_{\rm H}=R_{\rm H}/\rho$ with
increasing Ga content (Fig.\,3c top and bottom) strongly supports the picture of
vacancy filling because in a Ge-based clathrate electrons are expected to
scatter much more strongly from vacancies than from Ga.\vspace{0.4cm}

\noindent{\bf S4. Zintl-Klemm concept}

\noindent For the BCGG series a nearly temperature and sample independent charge
carrier concentration $n$ was found. According to the Zintl concept, Ba is
expected to donate 2 electrons to the host framework. In a first approximation,
the tetrahedrally coordinated host atoms can be considered to form covalent
bonds. The formal charge of a Ge/Ga/Cu atom within a type-I clathrate host
framework is expected to be close to 0/-1/-3. Therefore, one Ga atom
substituting for Ge (Cu) would change the charge carrier concentration by -1
(+2) electrons per formula unit. For the BCGG series, the variable amount of Ga
incorporated into the host framework leaves the charge carrier concentration
nearly unchanged. Therefore, a substitution of Ga for vacancies with a formal
charge of -1 is the most likely scenario.

\noindent{\bf S5. Specific heat analysis}

\noindent The phonon specific heat of type-I clathrates is typically described
by the sum of a Debye contribution and two Einstein contributions,
\begin{align*}\tag{S10}
C_{\rm p}= 9 N_{\rm D} R \left(\frac{T}{\Theta_{\rm D}}\right)^3 \int_0^{\Theta_{\rm D}/T} \frac{x^4 e^x}{\left(e^x-1\right)^2} dx + \sum_{\rm i=1}^{2}{p_{\rm i} N_{\rm Ei} R \left(\frac{\Theta_{\rm Ei}}{T}\right)^2 \frac{e^{\Theta_{\rm Ei}/T}}{\left(e^{\Theta_{\rm Ei}/T}-1\right)^2}}
\quad ,
\label{eq5}
\end{align*}
where $x=\hbar\omega/k_{\text{B}}T$ and $N_{\text{D}}$ and $N_{\text{Ei}}$ are
the numbers of Debye and Einstein oscillators per formula unit.
$\Theta_{\text{D}}$, $\Theta_{\text{Ei}}$, and $p_{\text{i}}$ are the Debye
temperature, Einstein temperatures, and the number of degrees of freedom related
to the \textit{i}-th vibrational mode of the guest atoms, respectively, and $R$
is the gas constant. In the fitting procedure we fixed $p_1=p_2=3$ and
$N_{\text{D}}+\sum{N_{\text{Ei}}}=54$. When presented as
$\frac{C_{\text{p}}}{T^3}$ vs $\log T$, Einstein contributions typically have a
bell-shaped appearance on top of a Debye background from which the corresponding
Einstein temperatures can be extracted by fitting with Eqn.\,\ref{eq5}.

Below 3.5\,K, the specific heat of all samples analyzed here has the form
$C_{\text{p}}/T =\gamma + \beta T^2$, which was used to extract the sound
velocity $v_{\rm s}$. The number of contributing atoms was assumed to be 54.
When inelastic neutron scattering data were available, $v_{\rm s}$ was extracted
from these using the relation $v_{\rm s} = \left[1/3\cdot((1/v_{\rm
L})^3+2\cdot(1/v_{\rm T})^3)\right]^{-1/3}$, where $v_{\rm T}$ and $v_{\rm L}$
are the initial slopes of the transverse and longitudinal acoustic phonon
branches, respectively. The agreement between both methods is satisfying.
\vspace{0.4cm}

\noindent{\bf S6. Theoretical calculation of the specific heat}

\noindent The vibrational properties of Ba$_8$Ga$_{16}$Ge$_{30}$ and Si
were computed with {\em ab initio} density functional theory (Methods). The
specific heat at constant volume was calculated from the phonon density of
states $D(\omega)$, shown in Fig.\,S1 insets, via
\begin{equation}\tag{S11}
C_{\rm V}(T) = \frac{\partial}{\partial T} \int_0^{\infty}d\omega D(\omega)\left[\frac{1}{e^{\frac{\hbar \omega}{k_{\rm B}T}}-1}\right] (\hbar \omega + \frac{1}{2})\quad.
\label{eq:CvDFT}
\end{equation}

\newpage

\noindent{\bf S7. Experimental determination of the thermal expansion}

\noindent The linear coefficient of thermal expansion $\alpha_{\rm{L}}(T)$ of
single crystalline Ba$_8$Ga$_{16}$Ge$_{30}$ was determined from the
experimentally revealed $\Delta l(T)/l$ data, with $\Delta l(T) = l(T) -
l(200$\,K), by numerical differentiation with respect to temperature. For that
purpose the $\Delta l(T)/l$ data were divided into equidistant intervals of
width $\Delta T = 0.13$\,K, in each of which the mean slope was determined from
linear regression. The experiment was carried out on a single crystal of length
$l = 1.6$\,mm along a principal axis of the cubic structure for which the volume
expansion coefficient $\alpha_{\rm{V}}$ is given by $\alpha_{\rm{V}} =
3\cdot\alpha_{\rm{L}}$. Measurements were performed from 4.5\,K up to 200\,K
upon increasing the temperature at a slow rate of 1.5\,K/h to ensure thermal
equilibrium. At low temperatures, the data are well described by
$\alpha_{\rm{V}}/T = a + b T^2$, with a negligibly small electronic contribution
$a$ (not shown). Thus, $\alpha_{\rm{L}}(T)$ is dominated by the phonon
contribution.\vspace{0.4cm}

\noindent{\bf S8. Phonon Kondo effect}

\noindent The Kondo problem has a 50-year-old history of highly active research.
It has been studied in settings ranging from bulk materials, to surfaces,
2-dimensional materials, mesoscopic systems, and ultracold atoms in optical
lattices. Originally it was formulated for free  conduction electrons
interacting with spin-1/2 local magnetic moments\cite{Kon64.1,Hew97.1,Col15.1}.
However, Kondo physics is a very broad phenomenon, with only a few fundamental
requirements, most notably the non-commutative scattering of an extended wave
off a localized entity with internal degrees of freedom.  In addition to
magnetic moments, the localized entities may take the form of
orbital\cite{Coq69.1,Noz80.1,Cox87.1}, charge\cite{Wil02.1}, and local
vibrational degrees of freedom\cite{Kat87.1,Cox98.1,Hot06.1,Hot07.1}.  The
extended wave component can also take various forms,  in addition to conduction
electrons in simple metals, it may also arise from  fermions with pseudo-gapped
electronic densities of states\cite{Wit90.1}, Dirac particles
\cite{Yan15.1,Che11.1}, and bosonic baths\cite{Dua04.1,Flo15.1}. To the best of
our knowledge Kondo physics has, however, never been considered in an entirely
phononic context.

In the following we give an experimentalist's sketch of how the basic physical
picture of a Kondo interaction between rattling modes and the transverse
acoustic phonons of a type-I clathrate, described in the main text, might be
translated into a Kondo-type Hamiltonian.

The two degenerate Einstein modes associated with the two rattling directions
($\alpha = 1, 2$) of the guest atoms in the large cages are described as
\begin{equation}\tag{S12}
H_{0} =  \sum_{\alpha= 1,2}\hbar\omega_{\rm{E}}
(b\dg_{\alpha}b_{\alpha}+\frac{1}{2}) + 
\frac{U}{2 } (b\dg_{1}b_1 + b\dg_{2}b_{2} - N_b)^{2}  \quad ,
\label{H0}
\end{equation}
where $b\dg_{\alpha}$ and $b_{\alpha}$ ($\alpha = 1,2$) are the boson creation
and annihilation operators, $N_b$ is an effective boson number, and $U$ is the
interaction energy.   We can absorb the first term into the second, writing 
\begin{equation}\tag{S13}
H_{0}= \frac{U}{2} (n_1 + n_2 - N_b^{\ast})^2  + \hbar\omega_{\rm{E}}(1+\frac{\hbar\omega_{\rm{E}}}{2U}+N_b^{\ast}) \quad ,
\end{equation}
where
\begin{equation}\tag{S14}
N_b^{\ast} = N_b - \frac{\hbar \omega_E}{U}
\end{equation}
and $b\dg_{\alpha}b_{\alpha} = n_{\alpha}$. Provided this quantity remains
positive, then the ground state energy is minimized at $n_{1}+n_{2 }=
N_b^{\ast}$, and the energy of the states with one more, or one less phonon is
$U/2$ higher. In this way, our local Hamiltonian has the effect of inducing a
``phonon blockade'', the rattler analog of the Coulomb blockade of the spin
Kondo effect, in which the states with $N_b^{\ast}$ phonons are degenerate, and
split off beneath all states with larger, or smaller numbers of bosons. The
Hamiltonian also ensures that there is an effective SU(2) degeneracy between the
``1'' and ``2'' bosons. However, it may well be that other formulations can be
found that represent a phonon Kondo effect more elegantly.

The hybridization of the Einstein modes with the acoustic modes is described by 
\begin{equation}\tag{S15}
H_{1}= \sum_{{\bf{q}},\alpha=1,2} V_{\bf{q}}(b_{\bf{q},\alpha}+b\dg_{-{\bf{q}},\alpha})(b_{\alpha}+b\dg_{\alpha}) \quad
\end{equation}
where $V_{\bf{q}}$ is the hybridization strength. It does not conserve the boson
number. This will produce additional Kondo terms in the Schrieffer-Wolff
transformation, in which the number of acoustic bosons changes by $\pm 2$. 

Finally, the acoustic modes themselves are described by
\begin{equation}\tag{S16}
H_{2}= \sum_{{\bf{q}},\alpha= 1,2}\hbar\omega_{{\bf{q}},\alpha} (b\dg_{{\bf{q}},\alpha}b_{{\bf{q}},\alpha}+\frac{1}{2}) \quad ,
\end{equation}
where $\bf{q}$ is the wave vector and $\alpha$ the polarization of an acoustic
phonon. The total Hamiltonian of the interacting acoustic and rattling phonon
system thus reads
\begin{equation}\tag{S17}
H= H_0 + H_1 + H_2 \quad ,
\end{equation}
which resembles an Anderson impurity model \cite{Hew97.1}.

To map this Hamiltonian directly to a spin Kondo Hamiltonian (Schrieffer-Wolff
transformation \cite{Hew97.1,Col15.1}) we note that the two-fold degeneracy of
the rattler modes ($\alpha = 1$,\,2) resembles a spin, described in a Schwinger
boson representation\cite{Rea83.1,Col87.1,Aro88.1} as
\begin{equation}\tag{S18}
S_{z}= \frac{1}{2}(b\dg_{1}b_1- b\dg_{2}b_{2}) \quad ,
\end{equation}
where
\begin{equation}\tag{S19}
S= \frac{1}{2}(b\dg_{1}b_1 + b\dg_{2}b_{2}) \quad .
\end{equation}
Interestingly, this means that the effective spin of the rattler is given by the
Bose-Einstein function
\begin{equation}\tag{S20}
2 S = N_b  = 1/ (e^{\hbar\omega_{\rm{E}}/(k_{\rm{B}}T)}-1)
\end{equation}
and will thus grow with $k_{\rm{B}}T/(\hbar\omega_{\rm{E}})$ at high
temperatures. This distinguishes the all phononic Kondo effect from the usual
spin Kondo effect in a simple metal, where the interaction does not disappear at
$T=0$. By contrast, it is similar to the Kondo effect in pseudo-gapped Fermi
systems \cite{Wit90.1} and in particular in fermionic systems with linear
dispersion like graphene and 3-dimensional Dirac semimetals. Here the electronic
density of states vanishes at the Fermi level and thus Kondo screening is
expected to disappear below the Kondo temperature\cite{Yan15.1}. Interestingly,
a $- \ln T$ dependence of the electrical resistivity is experimentally observed
in defect-graphene in a rather extended temperature range above some cutoff
temperature \cite{Che11.1}.

The acoustic phonon modes apparently have the ability to switch the polarization
of the rattler, which can be visualized as follows: If an acoustic phonon of
polarization 1 arrives at a cage with an atom rattling in polarization 2, it
will be absorbed in the cage in the form of a distortion of the cage along 1,
corresponding to an excited state (represented by the ``Hubbard'' term in
Eqn.\,\ref{H0}). This may lead to a change of the rattling direction ($2
\rightarrow 1$), and the subsequent emission of an acoustic phonon of changed
polarization ($1 \rightarrow 2$). Such a process is governed by the
``antiferromagnetic'' exchange interaction
\begin{equation}\tag{S21}
J \sim \frac{V^{2}}{U}
\end{equation}
where, for simplicity, we have assumed the hybridization to be wave vector
independent. This leads to the well-known Kondo term
\begin{equation}\tag{S22}
H_{\rm{K}} \sim J {\bf{s}}(0)\cdot 2{\bf{S}} \quad ,
\end{equation}
where $s(0)$ represents the effective pseudospin
\begin{equation}\tag{S23}
s(0) =  \phi\dg_{\alpha }\sigma_{\alpha \beta }\phi_{\beta },
\end{equation}
of the acoustic phonons at the position of the local pseudospin $2S$,
$\sigma_{\alpha\beta}$ are the Pauli matrices, and 
\begin{equation}\tag{S24}
\phi_{\alpha} = \sum_{\bf{q}} g_{{\bf{q}},\alpha}(b_{{\bf{q}},\alpha}+b\dg_{-{\bf{q}},\alpha})
\end{equation}
are the displacements of the acoustic modes that couple linearly to the rattler
modes.

To treat this or equivalent formulations of the all phononic Kondo effect we
believe to be realized in type-I clathrates and derive its physical properties
for a realistic phonon density of states as function of the coupling strength is
a task for future theoretical work. However, in analogy with related models (see
above), it seems likely that archetype Kondo features such as the observed
low-temperature ($-\ln T$-like) increase of the thermal resistance would indeed
result from such a model, at least in a limited temperature range.

\vspace{0.4cm}


\newpage

{\bf Table\,S1:} {\bf Materials obeying universal scaling.} Type-I clathrate single
crystals synthesized for this work (BCGG$x$, BGG) and materials used
for comparison. All intermetallic clathrates except the large-grain polycrystal
BAgG are single crystals. The other materials are polycrystals.\vspace{0.5cm}

\begin{center}
\begin{tabular}{lcr}
\textbf{Abbreviation} & \textbf{Composition}  & \textbf{Reference}\\
\hline
BCGG0.0    & Ba$_8$Cu$_{4.8}$Ge$_{40}\square_{1.2}$  &\\
BCGG0.2    & Ba$_8$Cu$_{4.8}$Ge$_{40}$Ga$_{0.2}\square_{1.0}$  &\\
BCGG0.5    & Ba$_8$Cu$_{4.8}$Ge$_{40}$Ga$_{0.5}\square_{0.7}$ &\\
BCGG1.0    & Ba$_8$Cu$_{4.8}$Ge$_{40}$Ga$_{1.0}\square_{0.2}$ &\\ 
BNG        & Ba$_8$Ni$_{3.5}$Ge$_{42.1}\square_{0.4}$ & \cite{Ngu10.1}\\
BAG        & Ba$_8$Au$_{5.25}$Ge$_{40.3}\square_{0.45}$ & \cite{Zha11.1,Tom15.1}\\
BAgG       & Ba$_8$Ag$_{4.1}$Ge$_{41.4}\square_{0.5}$ & \cite{Zei11.2}\\
La-BAS     & La$_{1.23}$Ba$_{6.99}$Au$_{5.91}$Si$_{39.87}$ & \cite{Pro13.1}\\
Ce-BAS     & Ce$_{1.06}$Ba$_{6.91}$Au$_{5.56}$Si$_{40.47}$ & \cite{Pro13.1}\\
BGG        & Ba$_8$Ga$_{16}$Ge$_{30}$ & \cite{Chr08.1,Sal01.1}\\
SGG        & Sr$_8$Ga$_{16}$Ge$_{30}$ &\cite{Sal01.1,Qiu04.1}\\
EGG        & Eu$_8$Ga$_{16}$Ge$_{30}$ &\cite{Sal01.1,Pas01.2}\\
BGSn       & Ba$_8$Ga$_{16}$Sn$_{30}$ (n-type) & \cite{Avi08.1,Ish12.1,Sai12.1}\\
Xe hydrate & Xe$\cdot 6.2$H$_2$O & \cite{Gab09.1,Han87.1,Tse01.1}\\
CMSS       & Cu$_{10.6}$Mn$_{1.4}$Sb$_{4}$S$_{13}$ & \cite{Che15.1}\\
\hline
\end{tabular}
\label{samples}
\end{center}


\newpage
\begin{figure}[!ht]
{\large\bf \hspace{-8cm} a \hspace{7.6cm} b}
\vspace{-0.1cm}\\
\subfigure{\includegraphics[width=0.5\textwidth]{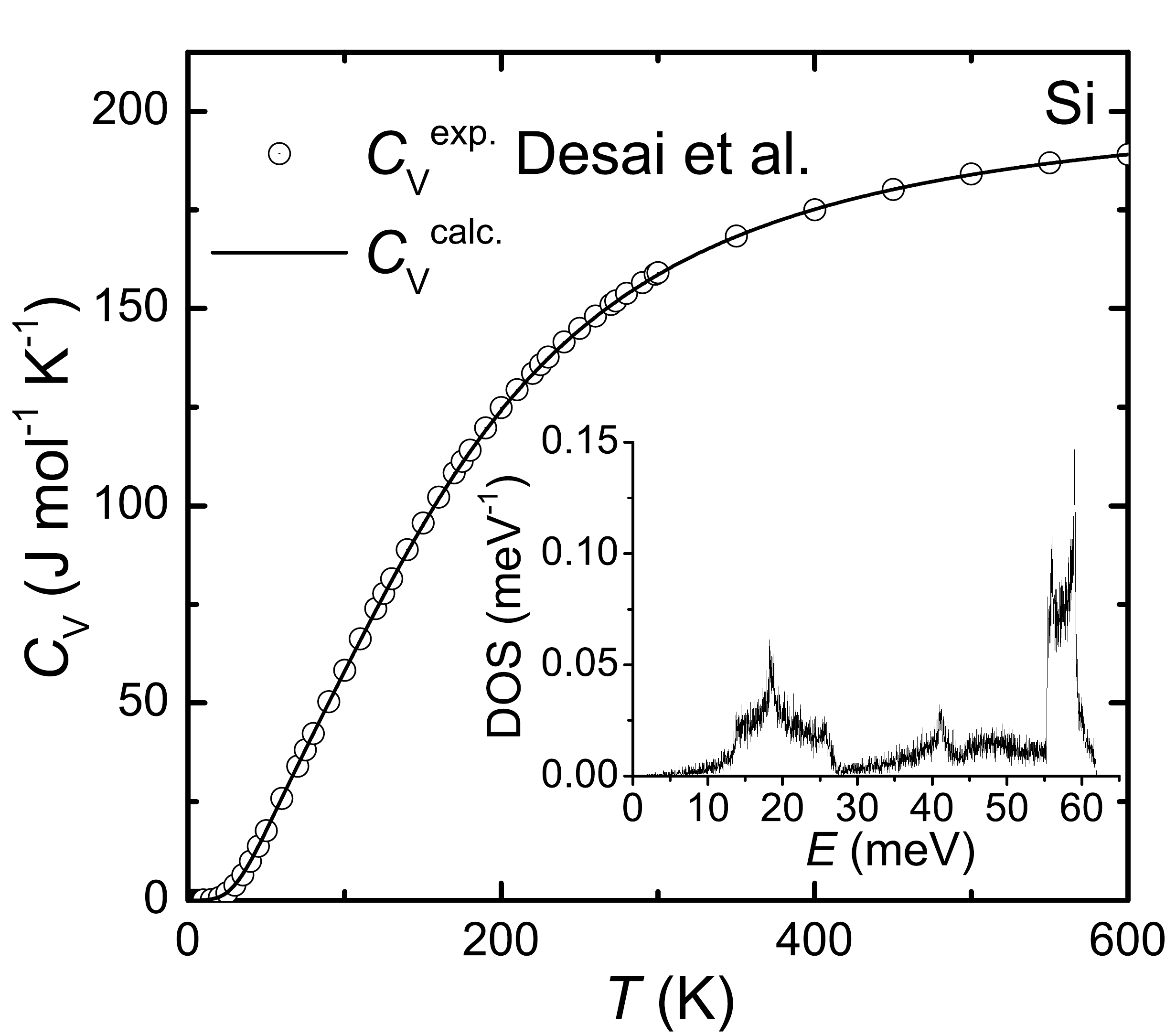}}\hfill
\subfigure{\includegraphics[width=0.5\textwidth]{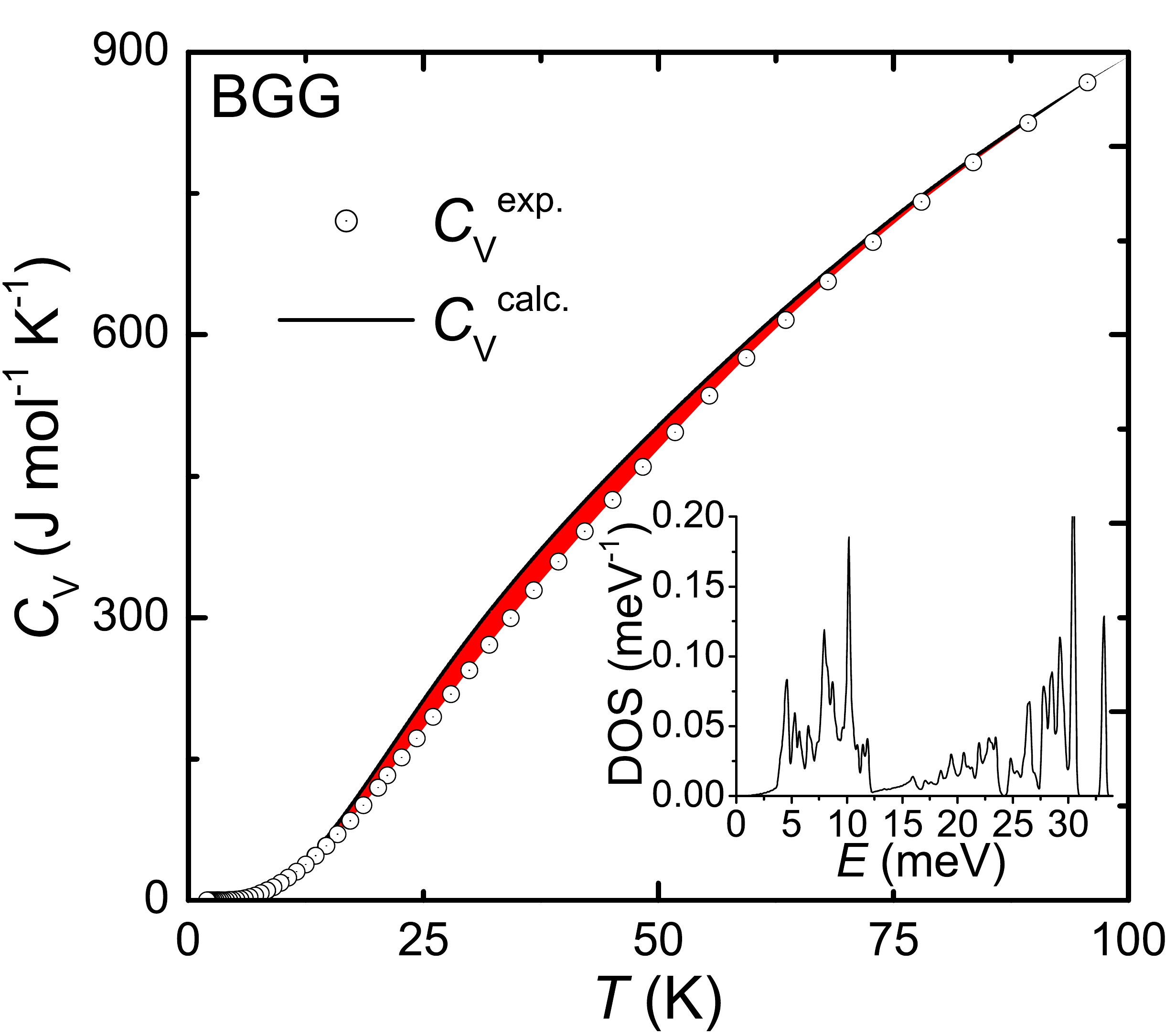}}
\end{figure}

{\bf Figure\,S1:} {\bf Specific heat of Si and Ba$_8$Ga$_{16}$Ge$_{30}$.}
Experimental (symbols) and calculated (lines) specific heat of (a) Si and (b)
Ba$_8$Ga$_{16}$Ge$_{30}$ (BGG) versus temperature. Whereas for Si, the
calculated specific heat is in good agreement with experimental
data\cite{Des85.1,Oka84.1}, for Ba$_8$Ga$_{16}$Ge$_{30}$ an extra contribution
(red area) is observed between 10 and 100\,K. As discussed in the main part,
this contribution can be attributed to the weakening of the correlation between
acoustic and optical phonon modes. The inset in both panels shows the \textit{ab
initio} phonon density of states (DOS) as a function of phonon energy, used
for calculating the specific heat (see Methods and Supplementary Information S6).


\newpage

\begin{figure}[!ht]
\vspace{-0.8cm}

{\large\bf \hspace{-8cm} a \hspace{7.6cm} b}
\vspace{-0.1cm}\\
\subfigure{\includegraphics[width=0.48\textwidth]{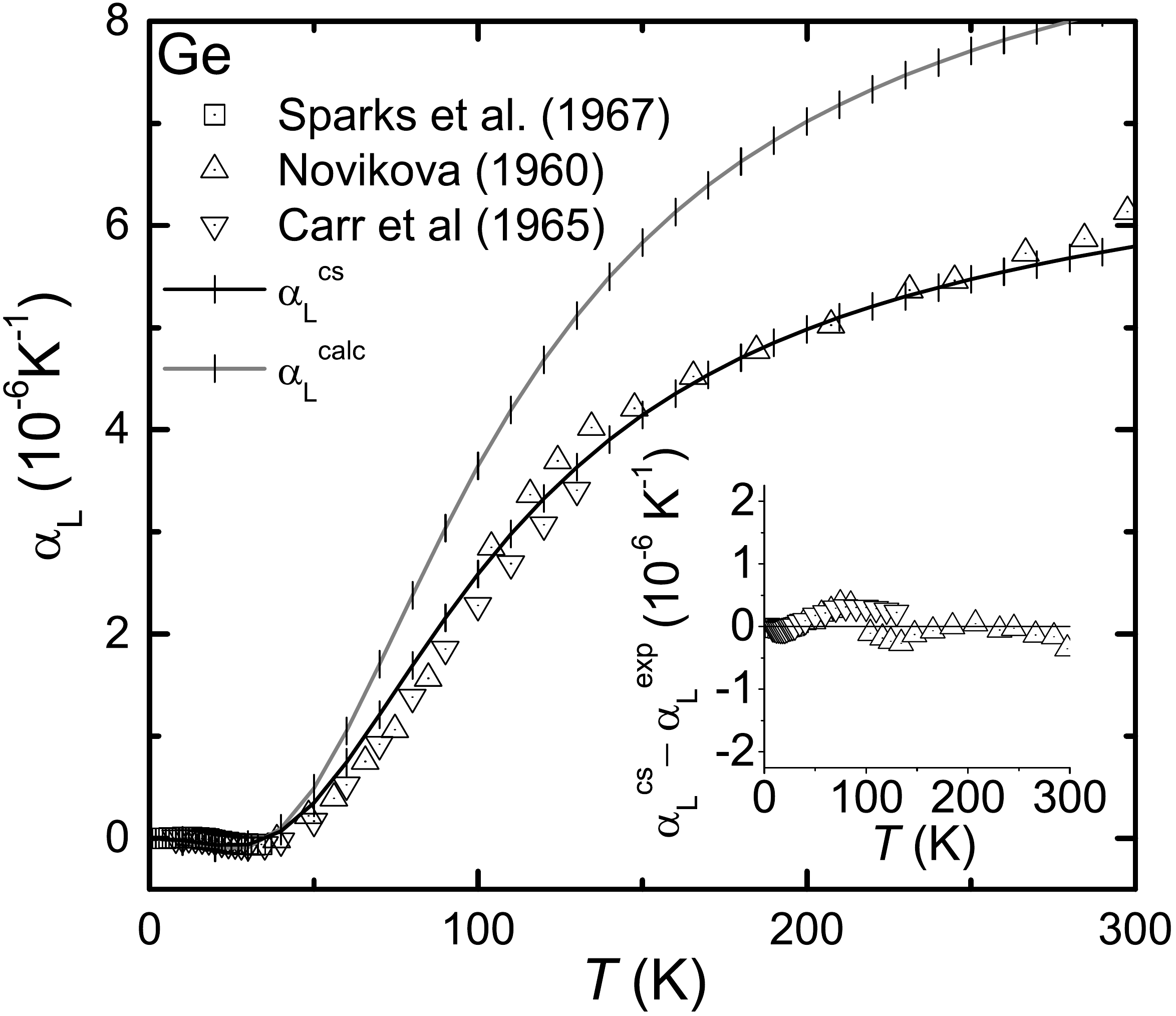}}\hfill
\subfigure{\includegraphics[width=0.48\textwidth]{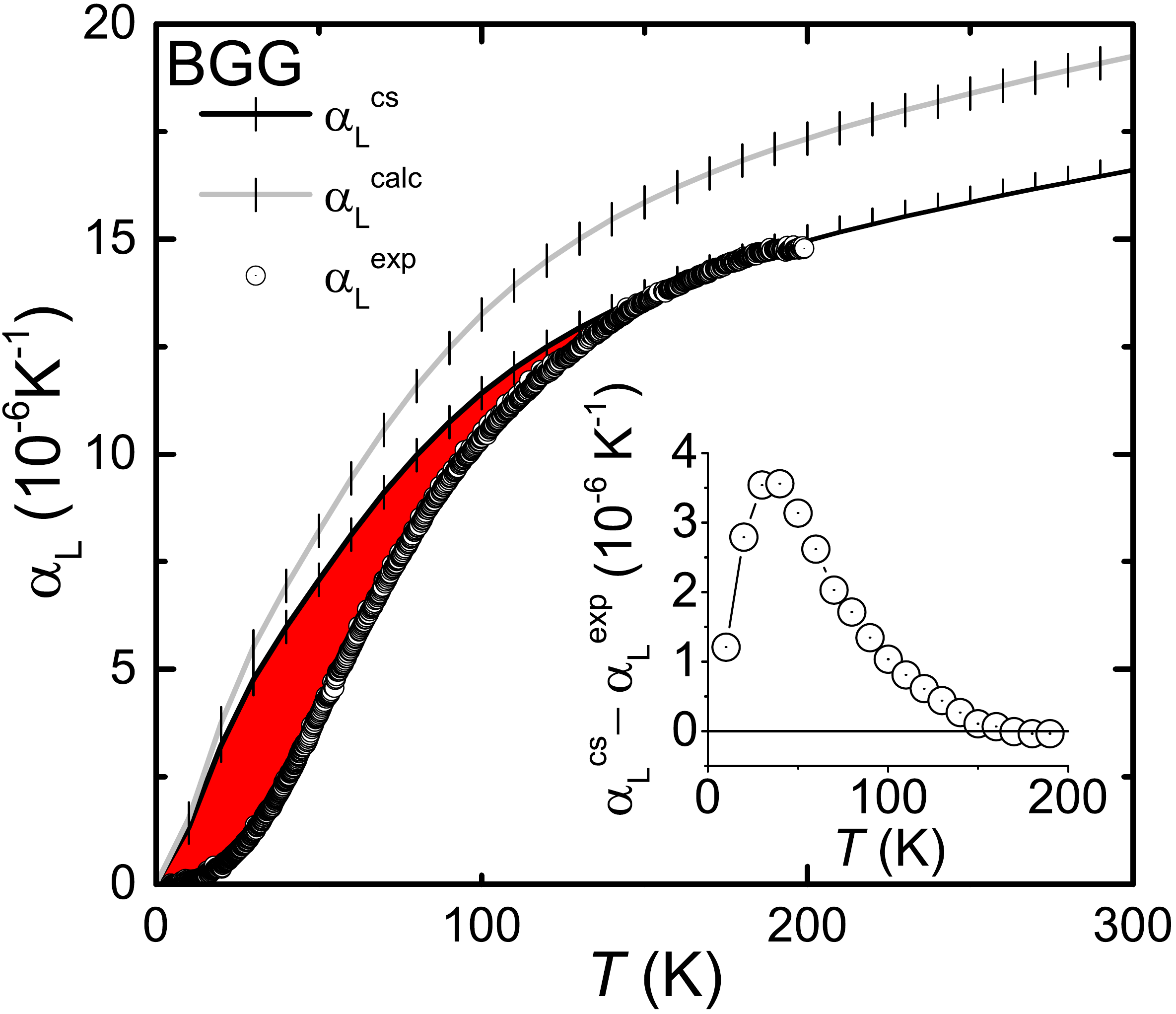}}\\[-0.2cm]
{\large\bf \hspace{-8.2cm} c}
\vspace{-0.3cm}\\

\subfigure{\includegraphics[width=0.48\textwidth]{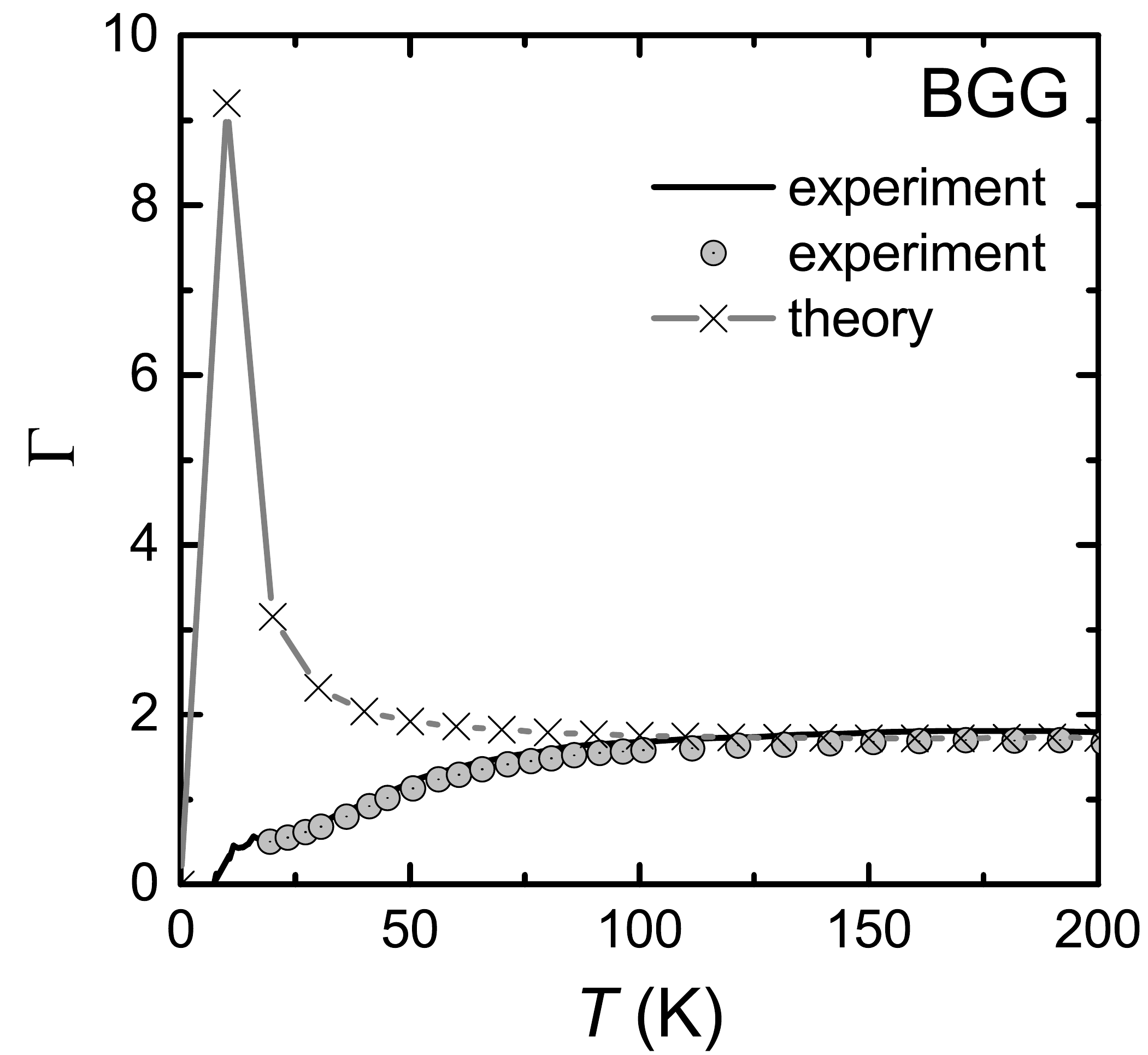}}
\end{figure}
\vspace{-0.8cm}

{\bf Figure\,S2:} {\bf Thermal expansion of Ge and Ba$_8$Ga$_{16}$Ge$_{30}$,
\SBP{and Gr\"{u}neisen parameter of Ba$_8$Ga$_{16}$Ge$_{30}$.}} Experimental
($\alpha_{\rm L}^{\rm exp}$, symbols), calculated ($\alpha_{\rm L}^{\rm calc}$,
gray lines), and rescaled ($\alpha_{\rm L}^{\rm cs}$, black lines) thermal
expansion of (a) Ge and (b) Ba$_8$Ga$_{16}$Ge$_{30}$ (BGG) versus temperature.
Whereas for Ge, the calculated (and rescaled) thermal expansion is in good
agreement with experimental data\cite{Car65.1,Spa67.1,Nov60.1}, for
Ba$_8$Ga$_{16}$Ge$_{30}$ an extra contribution (red area) is observed below
150\,K. The difference between theory and experiment is plotted in the two
insets. \SBP{(c) Comparison of the experimental Gr\"{u}neisen parameter $\Gamma
= B \alpha_{\rm V}/C_{\rm V}$ of Ba$_8$Ga$_{16}$Ge$_{30}$, using either a
temperature-independent bulk modulus $B$ of 65\,GPa (black solid line) or the
temperature-dependent one of Eu$_8$Ga$_{16}$Ge$_{30}$ (Ref.\citenum{Zer04.1},
gray circles), and the theoretical Gr\"{u}neisen parameter determined from our
{\em{ab initio}} lattice dynamics calculations (crosses and gray line). $C_{\rm
V}$ is the specific heat at constant volume, $\alpha_{\rm V}$ is the volumetric
thermal expansion coefficient.}


\newpage
\begin{figure}[!ht]
{\large\bf \hspace{-8cm} a \hspace{7.6cm} b}
\vspace{-0.1cm}\\
\subfigure{\includegraphics[width=0.5\textwidth]{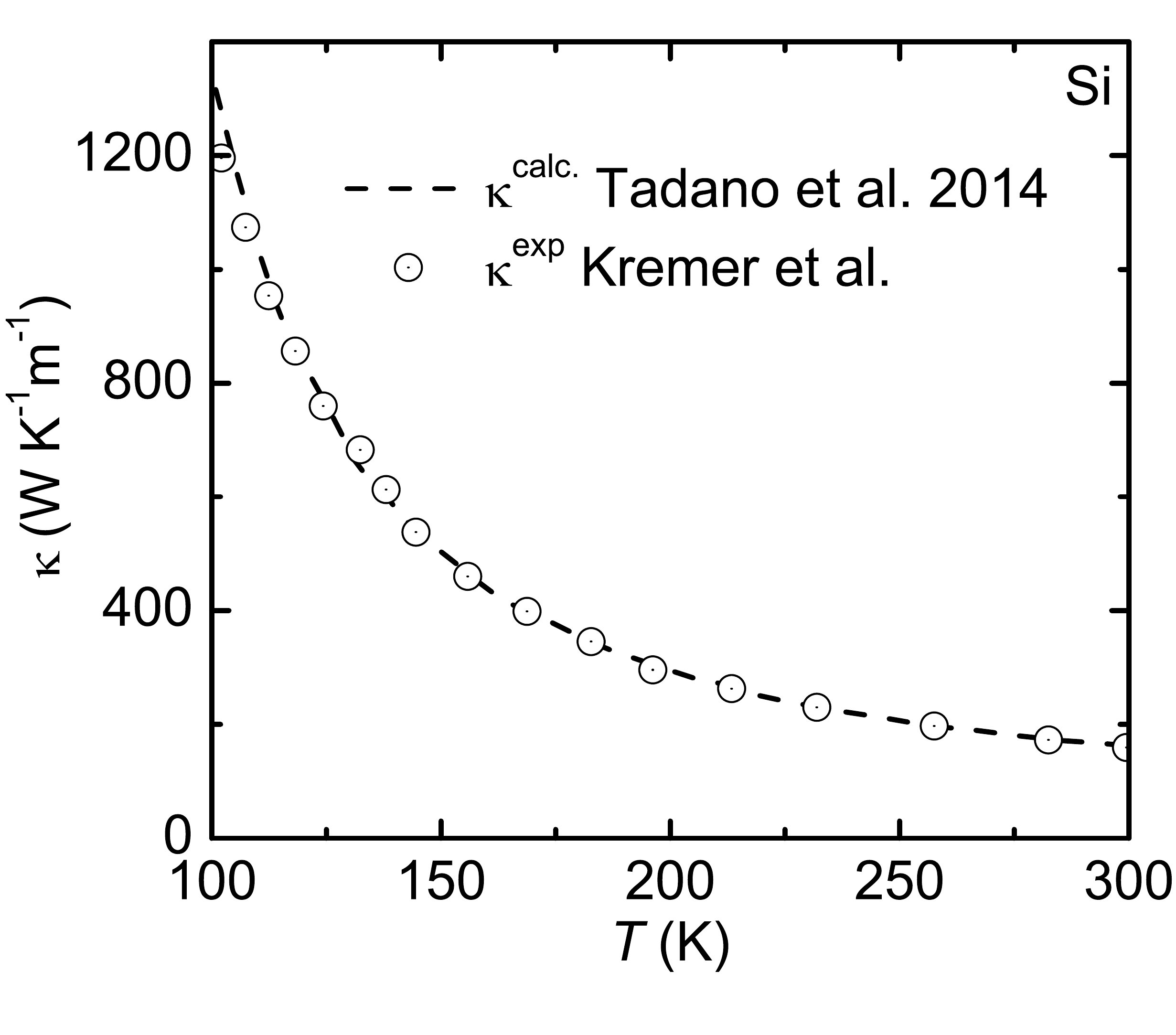}}\hfill
\subfigure{\includegraphics[width=0.5\textwidth]{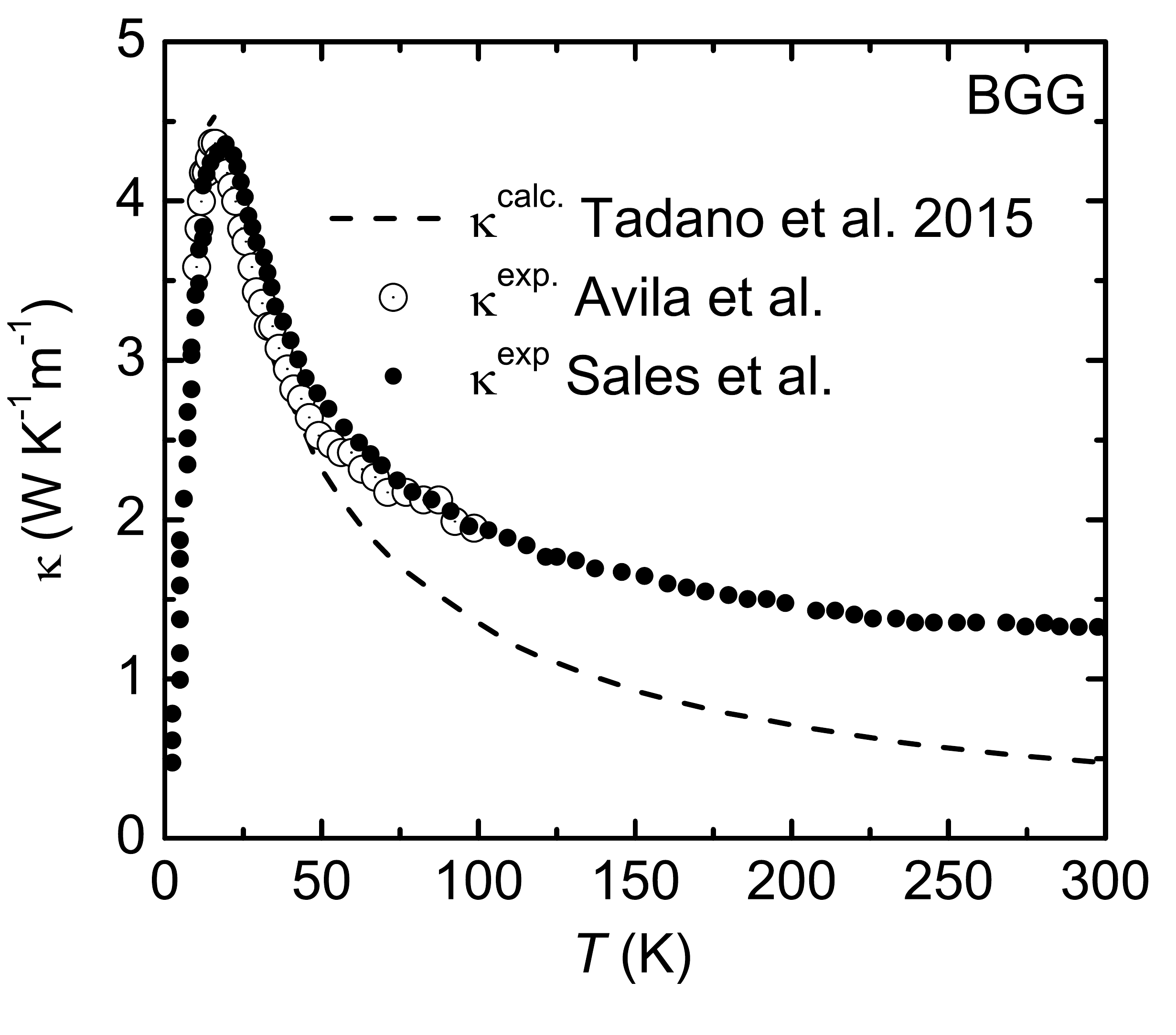}}
\end{figure}

{\bf Figure\,S3:} {\bf Thermal conductivity of Si and Ba$_8$Ga$_{16}$Ge$_{30}$.}
Experimental (symbols) and calculated (dashed lines) thermal conductivity of (a)
Si and (b) Ba$_8$Ga$_{16}$Ge$_{30}$ (BGG) versus temperature. While for Si, the
calculated thermal conductivity\cite{Tad14.1} is in good agreement with
experimental data\cite{Kre04.1}, for Ba$_8$Ga$_{16}$Ge$_{30}$ the calculated
thermal conductivity\cite{Tad15.1} undershoots the experimental
data\cite{Avi06.1,Sal01.1} significantly above 50\,K. As discussed in the main
part, the inverse thermal conductivity difference shows approximate $-\ln T$
behaviour, the hallmark of incoherent Kondo scattering in spin Kondo systems.


\newpage
\begin{figure}[!ht]
\vspace{-0.1cm}
\subfigure{\includegraphics[width=0.9\textwidth]{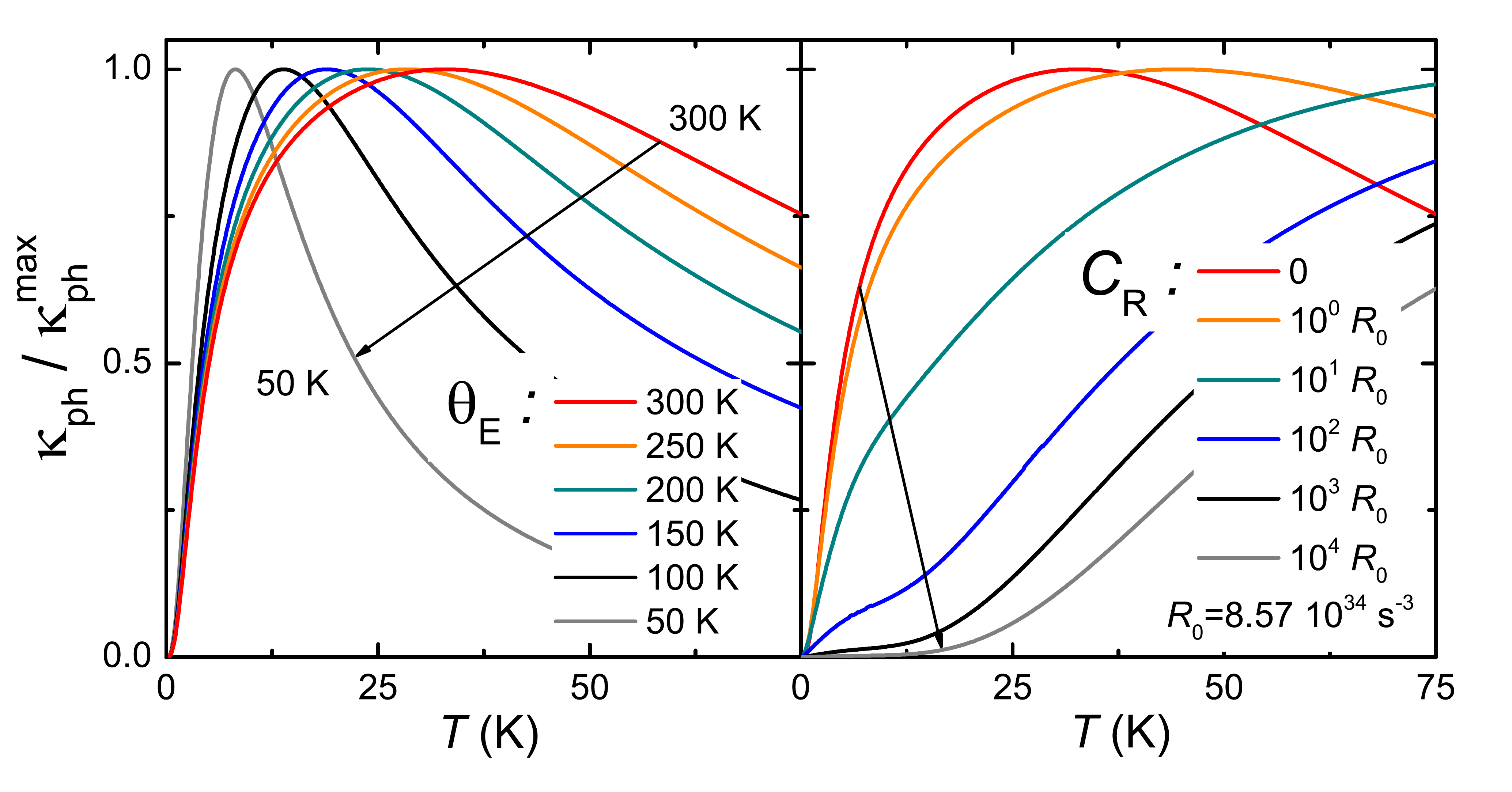}}
\end{figure}

{\bf Figure\,S4:} {\bf Enhanced Umklapp versus resonance scattering.} 
Temperature dependent phonon thermal conductivity, normalized to its maximum,
calculated using a modified Callaway model for various Einstein temperatures
$\Theta_{\rm E}$ (left, replotted from Fig.\,1d left) and for the
$\Theta_{\rm E} = 300$\,K curve modified by resonance scattering\cite{Poh62.1}
according to $\tau_{\rm R}^{-1}= \frac{C_{\rm R}
\omega^2}{\left(\omega_0^2-\omega^2\right)^2}$, where $\omega_0$ is the circular
frequency of the lowest-lying optical mode, with different scattering levels
$C_{\rm R}$ (right). Whereas enhanced Umklapp scattering leads to a sharpening
of the phonon thermal conductivity maximum at low temperatures and a suppression
of the thermal conductivity in a wide temperature range above the maximum,
resonance scattering leads to an additional shoulder and a reduction of the
thermal conductivity at low temperatures. Resonance scattering can thus be ruled
out as an important scattering channel for the type-I clathrates studied here.

\newpage
\noindent{\bf \large Additional References}\\